\documentclass{emulateapj}
\pdfoutput=1
\usepackage{graphicx}
\usepackage{epsfig,hyperref}
\usepackage{amssymb,amsmath}
\usepackage{array}

\singlespace

\newcommand{\map}{{\sl WMAP}}
\newcommand{\planck}{{\it Planck}}

\newcommand{\ba}{\begin{eqnarray}}
\newcommand{\ea}{\end{eqnarray}}

\newcommand{\uKrs}   {\mbox{$\mu{\rm K}\sqrt{\rm s}$}}

\newcommand  \beq    {\begin{equation}}

\newcommand  \eeq    {\end{equation}}
\newcommand  \gtsim  {\lower.5ex\hbox{$\; \buildrel > \over \sim \;$}}
\newcommand  \ltsim  {\lower.5ex\hbox{$\; \buildrel < \over \sim \;$}}

\newcommand{\LCDM}   {$\Lambda$CDM}
\newcommand{\jcap}   {JCAP}

\def\apfreq{146\,GHz}  
\def\pbear{{\textsc{Polarbear}}}
\def\taua{\mbox{Tau A}} 

\usepackage{color}
\definecolor{orange}{rgb}{1,0.3,0}

\shortauthors{}
\shorttitle{ACTPol polarization results}
\slugcomment{Draft: \today}

\begin{document}

\title{The Atacama Cosmology Telescope: CMB Polarization
 at $200<\ell<9000$}

\author{
Sigurd~Naess\altaffilmark{1},
Matthew~Hasselfield\altaffilmark{2,3},
Jeff~McMahon\altaffilmark{4},
Michael~D.~Niemack\altaffilmark{5},
Graeme~E.~Addison\altaffilmark{3},
Peter~A.~R.~Ade\altaffilmark{6},
Rupert~Allison\altaffilmark{1},
Mandana~Amiri\altaffilmark{3},
Nick~Battaglia\altaffilmark{7},
James~A.~Beall\altaffilmark{8},
Francesco~de~Bernardis\altaffilmark{5},
J~Richard~Bond\altaffilmark{9},
Joe~Britton\altaffilmark{8},
Erminia~Calabrese\altaffilmark{1},
Hsiao-mei~Cho\altaffilmark{8},
Kevin~Coughlin\altaffilmark{4},
Devin~Crichton\altaffilmark{10},
Sudeep~Das\altaffilmark{11},
Rahul~Datta\altaffilmark{4},
Mark~J.~Devlin\altaffilmark{12},
Simon~R.~Dicker\altaffilmark{12},
Joanna~Dunkley\altaffilmark{1},
Rolando~D\"{u}nner\altaffilmark{13},
Joseph~W.~Fowler\altaffilmark{8},
Anna~E.~Fox\altaffilmark{8},
Patricio~Gallardo\altaffilmark{5,13},
Emily~Grace\altaffilmark{14},
Megan~Gralla\altaffilmark{10},
Amir~Hajian\altaffilmark{9},
Mark~Halpern\altaffilmark{3},
Shawn~Henderson\altaffilmark{5},
J.~Colin~Hill\altaffilmark{2},
Gene~C.~Hilton\altaffilmark{8},
Matt~Hilton\altaffilmark{15},
Adam~D.~Hincks\altaffilmark{3},
Ren\'ee~Hlozek\altaffilmark{2},
Patty~Ho\altaffilmark{14},
Johannes~Hubmayr\altaffilmark{8},
Kevin~M.~Huffenberger\altaffilmark{16},
John~P.~Hughes\altaffilmark{17},
Leopoldo~Infante\altaffilmark{13},
Kent~Irwin\altaffilmark{18},
Rebecca~Jackson\altaffilmark{4,19},
Simon~Muya~Kasanda\altaffilmark{15,20},
Jeff~Klein\altaffilmark{12},
Brian~Koopman\altaffilmark{5},
Arthur~Kosowsky\altaffilmark{21},
Dale~Li\altaffilmark{8},
Thibaut~Louis\altaffilmark{1},
Marius~Lungu\altaffilmark{12},
Mathew~Madhavacheril\altaffilmark{22},
Tobias~A.~Marriage\altaffilmark{10},
Lo\"ic~Maurin\altaffilmark{13},
Felipe~Menanteau\altaffilmark{23,24},
Kavilan~Moodley\altaffilmark{15},
Charles~Munson\altaffilmark{4},
Laura~Newburgh\altaffilmark{14},
John~Nibarger\altaffilmark{8},
Michael~R.~Nolta\altaffilmark{9},
Lyman~A.~Page\altaffilmark{14},
Christine~Pappas\altaffilmark{14},
Bruce~Partridge\altaffilmark{25},
Felipe~Rojas\altaffilmark{13,26},
Benjamin~L.~Schmitt\altaffilmark{12},
Neelima~Sehgal\altaffilmark{22},
Blake~D.~Sherwin\altaffilmark{27},
Jon~Sievers\altaffilmark{20,9},
Sara~Simon\altaffilmark{14},
David~N.~Spergel\altaffilmark{2},
Suzanne~T.~Staggs\altaffilmark{14},
Eric~R.~Switzer\altaffilmark{28,9},
Robert~Thornton\altaffilmark{29,12},
Hy~Trac\altaffilmark{7},
Carole~Tucker\altaffilmark{6},
Masao~Uehara\altaffilmark{26},
Alexander~Van~Engelen\altaffilmark{22},
Jonathan~T.~Ward\altaffilmark{12},
Edward~J.~Wollack\altaffilmark{28}
}
\altaffiltext{1}{Sub-Department of Astrophysics, University of Oxford, Keble Road, Oxford, UK OX1 3RH}
\altaffiltext{2}{Department of Astrophysical Sciences, Peyton Hall, 
Princeton University, Princeton, NJ USA 08544}
\altaffiltext{3}{Department of Physics and Astronomy, University of
British Columbia, Vancouver, BC, Canada V6T 1Z4}
\altaffiltext{4}{Department of Physics, University of Michigan, Ann Arbor, USA 48103}
\altaffiltext{5}{Department of Physics, Cornell University, Ithaca, NY, USA 14853}
\altaffiltext{6}{School of Physics and Astronomy, Cardiff University, The Parade, 
Cardiff, Wales, UK CF24 3AA}
\altaffiltext{7}{McWilliams Center for Cosmology, Carnegie Mellon University, Department of Physics, 5000 Forbes Ave., Pittsburgh PA, USA, 15213}
\altaffiltext{8}{NIST Quantum Devices Group, 325
Broadway Mailcode 817.03, Boulder, CO, USA 80305}
\altaffiltext{9}{Canadian Institute for Theoretical Astrophysics, University of
Toronto, Toronto, ON, Canada M5S 3H8}
\altaffiltext{10}{Dept. of Physics and Astronomy, The Johns Hopkins University, 3400 N. Charles St., Baltimore, MD, USA 21218-2686}
\altaffiltext{11}{Department of High Energy Physics, Argonne National Laboratory, 9700 S Cass Ave, Lemont, IL USA 60439}
\altaffiltext{12}{Department of Physics and Astronomy, University of
Pennsylvania, 209 South 33rd Street, Philadelphia, PA, USA 19104}
\altaffiltext{13}{Departamento de Astronom{\'{i}}a y Astrof{\'{i}}sica, Pontific\'{i}a Universidad Cat\'{o}lica,
Casilla 306, Santiago 22, Chile}
\altaffiltext{14}{Joseph Henry Laboratories of Physics, Jadwin Hall,
Princeton University, Princeton, NJ, USA 08544}
\altaffiltext{15}{Astrophysics and Cosmology Research Unit, School of Mathematics, Statistics and Computer Science, University of KwaZulu-Natal, Durban 4041, South Africa}
\altaffiltext{16}{Department of Physics, Florida State University, Tallahassee FL, USA 32306}
\altaffiltext{17}{Department of Physics and Astronomy, Rutgers, 
The State University of New Jersey, Piscataway, NJ USA 08854-8019}
\altaffiltext{18}{Department of Physics, Stanford University, Stanford, CA, 
USA 94305-4085}
\altaffiltext{19}{School of Earth and Space Exploration, Arizona State University, Tempe, AZ, USA 85287}
\altaffiltext{20}{Astrophysics and Cosmology Research Unit, School of Chemistry and Physics, University of KwaZulu-Natal, Durban 4041, South Africa}
\altaffiltext{21}{Department of Physics and Astronomy, University of Pittsburgh, 
Pittsburgh, PA, USA 15260}
\altaffiltext{22}{Physics and Astronomy Department, Stony Brook University, Stony Brook, NY USA 11794}
\altaffiltext{23}{National Center for Supercomputing Applications (NCSA), University of Illinois at Urbana-Champaign, 1205 W. Clark St., Urbana, IL, USA, 61801}
\altaffiltext{24}{Department of Astronomy, University of Illinois at Urbana-Champaign, W. Green Street, Urbana, IL, USA, 61801}
\altaffiltext{25}{Department of Physics and Astronomy, Haverford College,
Haverford, PA, USA 19041}
\altaffiltext{26}{Sociedad Radiosky Asesor\'{i}as de Ingenier\'{i}a Limitada Lincoy\'{a}n 54,
Depto 805 Concepci\'{o}n, Chile}
\altaffiltext{27}{Berkeley Center for Cosmological Physics, LBL and
Department of Physics, University of California, Berkeley, CA, USA 94720}
\altaffiltext{28}{NASA/Goddard Space Flight Center, Greenbelt, MD, USA 20771}
\altaffiltext{29}{Department of Physics , West Chester University 
of Pennsylvania, West Chester, PA, USA 19383}

\begin{abstract}
We report on measurements of the cosmic microwave background (CMB) and celestial
polarization at \apfreq{} made with the Atacama Cosmology Telescope Polarimeter
(ACTPol) in its first three months of observing. Four regions of sky covering a
total of 270 square degrees were mapped with an angular resolution of $1.3'$.
The map noise levels in the four regions are between 11 and 17 $\mu$K-arcmin.
We present TT, TE, EE, TB, EB, and BB power spectra from three of these
regions. The observed E-mode polarization power spectrum, displaying six
acoustic peaks in the range $200<\ell<3000$, is an excellent fit to the
prediction of the best-fit cosmological models from WMAP9+ACT and \planck\
data. The polarization power spectrum, which mainly reflects primordial plasma
velocity perturbations, provides an independent determination of cosmological
parameters consistent with those based on the temperature power spectrum, which
results mostly from primordial density perturbations. We find that without
masking any point sources in the EE data at $\ell<9000$, the Poisson tail of
the EE power spectrum due to polarized point sources has an amplitude less than
$2.4$ $\mu$K$^2$ at $\ell = 3000$ at 95\% confidence. Finally, we report that
the Crab Nebula, an important polarization calibration source at microwave
frequencies, has 8.7\% polarization with an angle of $150.7^\circ \pm
0.6^\circ$  when smoothed with a $5'$ Gaussian beam.

\end{abstract}

\section{Introduction}
\label{sec:intro}
\setcounter{footnote}{0} 

The standard $\Lambda$CDM model of cosmology is now well established \citep[e.g.,][]{spergel/etal/2003,planck_params/2013}.
The model is geometrically flat and dark-energy dominated, with baryons comprising about 15\% of the total matter density. The initial fluctuations are adiabatic, Gaussian, and nearly scale-free.
The free parameters of the model have
been determined to high precision, largely from measurements
of the cosmic microwave background (CMB) temperature anisotropies on
angular scales from the full sky down to a few arcminutes \citep{hinshaw/etal/2013,planck_params/2013,sievers/etal/2013,hou/etal/2014}. Large-scale structure measurements of the baryon
acoustic oscillation scale \citep[e.g.,][]{anderson/etal/2014} and direct measurements of the expansion rate and acceleration from type-Ia
supernova observations \citep[e.g.,][]{suzuki/etal/2012,betoule/etal/2014} also provide constraints.
While the model is in generally good agreement with all cosmological probes, there are some mild tensions at the $2\sigma$ to $3\sigma$ level \citep[e.g.,][]{planck_params/2013} and possible hints of ``anomalies'' \citep[e.g.,][]{planck_isot/2013,bennett/etal/2011,copi/etal/2010}. Further precise tests of the model are clearly needed.

The polarization of the CMB provides a complementary source of information
and also probes cosmological physics beyond what can be obtained
with temperature alone. The E-mode polarization power spectrum
\citep{kamionkowski/etal/1997,zaldarriaga/seljak/1997} arises primarily
from the motions of the primordial plasma at the
epoch of last scattering (redshift $z=1100$),
with a contribution at large angular scales from perturbations
at the reionization epoch ($z\approx 10$).  Precision measurements
of the acoustic features in the E-mode polarization power spectrum
provide a non-trivial confirmation of the acoustic oscillations in
the early universe seen in temperature data. \citet{rocha/etal/2004} and \citet{galli/etal/2014} point out that with the CMB E-mode polarization alone one may
place stronger constraints on cosmological parameters
than with the temperature anisotropy.
More importantly,  the combination of temperature and polarization
data constrain
a range of physical effects beyond the standard model.
For example, primordial isocurvature perturbations alter the phase of the
oscillations in polarization relative to temperature compared
to pure adiabatic perturbations \citep[e.g.,][]{bond/efstathiou/1987,bucher/etal/2004,mactavish/etal/2006,sievers/etal/2007}.
The additional sensitivity to the
standard cosmological model provides increased ability to probe
neutrino properties, early dark energy, and time variation of
fundamental constants. Comparison of temperature and
polarization data also probes cosmological effects which affect the two
in different ways, such as the kinematic Sunyaev-Zeldovich effect \citep{calabrese/etal/2014}.

Polarization observations can also characterize
the B-mode fluctuations, which are not generated by the dominant
primordial density perturbations. B-mode polarization
contains a signal from gravitational lensing by all structure along
the line of sight, for multipoles $\ell > 200$ \citep{zaldarriaga/seljak/1998}.
The B-mode fluctuations also reflect
other new physical effects, such as cosmic
birefringence due to magnetic fields \citep{kosowsky/etal/2005} or photon couplings beyond
the standard model \citep{lue/etal/1999}.
A definitive detection of primordial B-mode polarization at $\ell < 200$ from
tensor perturbations generated by inflation will test the fundamental
nature of the gravitational
force and probe energy scales well beyond
terrestrial experiments.

Although \map\ has made the only published measurement of
large-angle E-mode polarization at $\ell< 15$,
measurements
of polarization with $\ell > 15$ have steadily improved.
The \planck\ team has shown
excellent visual agreement between their
best-fit $\Lambda$CDM model and their TE and EE polarization spectra for
$\ell \geq 100$ but has not yet quantified the agreement \citep[][Figure 11]{planck_params/2013}. On
larger angular scales, $30 < \ell < 1000$,
the CAPMAP \citep{bischoff/etal/2008}, QUAD \citep{brown/etal/2009}, QUIET \citep{quiet-w/2012},
and BICEP \citep{bicep/2013}
teams have shown, along with \map\ \citep{hinshaw/etal/2013},
that the predicted E-mode signal is in
quantitative agreement with the $\Lambda$CDM prediction. In
addition, they placed limits on primordial and lensing
B-modes.
Through cross correlating the EB-reconstructed lensing signal
with the Herschel-SPIRE maps, SPT \citep{hanson/etal/2013} demonstrated the presence of lensing B-mode
polarization at $7.7\sigma$. In a
similar cross-correlation analysis, the \pbear{} team found $2.3\sigma$ evidence
\citep{pbear-herschel/2013} for lensing
B-modes and also demonstrated the presence
of lensed B-mode polarization through the EEEB and
EBEB four-point functions at $4.2\sigma$ \citep{pbear-eebb/2014}.
Following that, \pbear{}
released their measurements of the TT, TE, and EE power
spectra, $2\sigma$ evidence
of non-zero BB power, and limits on the TB and EB spectra
\citep{pbear-eebb/2014}. Most recently,
the BICEP2 team released their $7\sigma$ detection of degree angular-scale B-mode polarization along with a suite of
related spectra \citep{bicep2a/2014}.

This paper is the Atacama Cosmology Telescope (ACT) collaboration's
first step in measuring
CMB polarization and is organized as follows. In \S2 we
introduce the salient features of the instrument and then
in \S3 describe the observations and data reduction. In \S4
we present our power spectra measurements and interpretation, and conclude in \S5. For power
spectra, we use the {\it Planck} notation: ${\cal D}^{\rm{XY}}_{\ell} = \ell(\ell+1) C^{\rm XY}_\ell / 2\pi$
where ${\rm XY} \in {{\rm TT,\,TE,\, TB,\, EE,\, EB,\, BB}}$.
We do not consider circular polarization \citep[e.g.,][]{alexander/etal/2009,cooray/etal/2003}.
The maps are made in J2000 equatorial coordinates.  We adopt the
HEALPIX \citep{gorski/etal/2005} convention for Stokes parameters Q
and U.  Polarization position angles respect the IAU
convention \citep[see, e.g.,][]{hamaker/bregman/1996}, increasing from
North towards East, and thus are computed as $\gamma_p =
(1/2) \text{arg}(Q - iU)$.  We note that these are the same
conventions adopted by \planck{} \citep[see, e.g., Section 2.1
of][]{planck_dust/2014}, although \planck\ uses Galactic
coordinates.

\begin{figure*}
	\includegraphics[width=\textwidth]{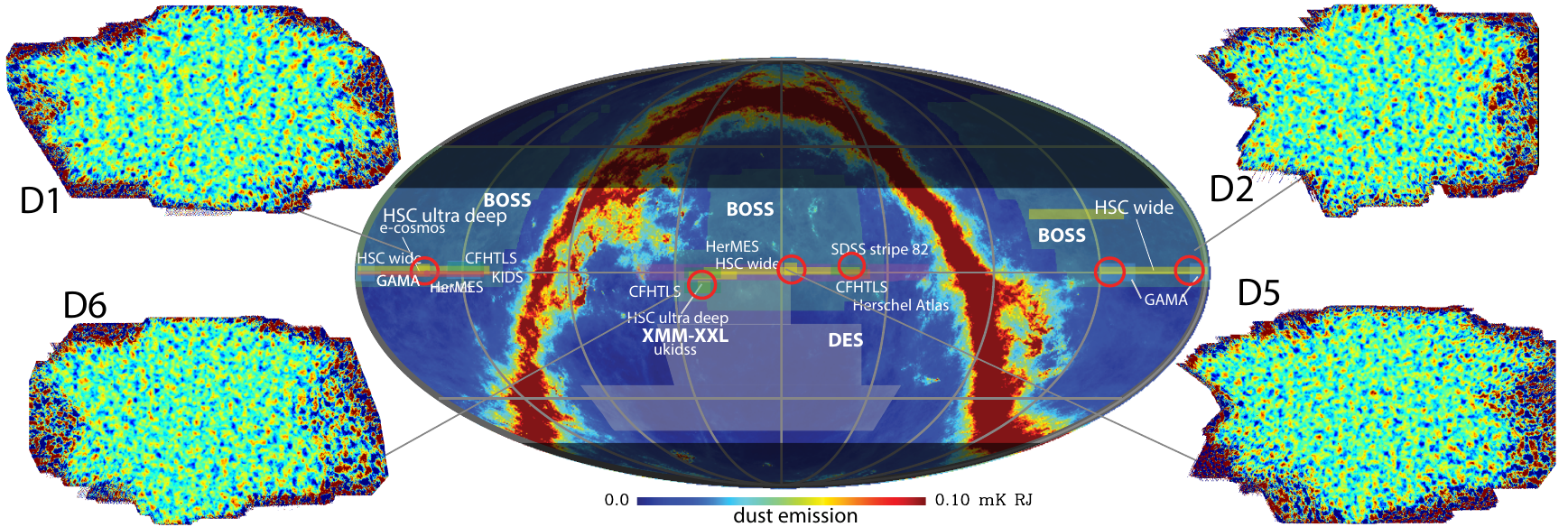}
\caption{ACTPol maps and overlapping surveys. The maps have been filtered to
emphasize $\ell>300$. The power spectra are obtained with only the high S/N
region of each map. Going from left to right across the equator, the
red circles indicate patches D1, D6, D5,
D4, D3 and D2 (the first ACTPol season focused on D1, D2, D5 and D6).  More than half the sky, as indicated by the light colored area, is
accessible to ACTPol. Overlapping surveys include SDSS \citep{sdss/2014},
BOSS\citep{boss/2014}, CFHTLS \citep{cfhtls/2013}, XMM-XXL
\citep{xmm/2014}, Herschel (HerMES \& HeLMS, \citet{oliver/etal/2012,
viero/etal/2014}), HSC \citep{hsc/2014}, DES \citep{des/2014}, GAMA
\citep{gama/2009}, and KiDS \citep{kids/2013}.}
\label{fig:maps}
\end{figure*}

\section{The Instrument}
\label{sec:inst}
ACT is located at latitude 22$^\circ$57$'$31$''$S and longitude
67$^\circ$47$'$15$''$W at an altitude of 5190\,m in Parque
Astron\'omico Atacama in northern Chile.  The 6\,m primary mirror
leads to arcminute resolution at millimeter wavelengths.  A first-generation receiver, the Millimeter Bolometric Array Camera
\citep[MBAC;][]{fowler/etal/2007, swetz/etal/2011}, observed from
Sept. 2007 through Dec. 2010. It was decommissioned and replaced by
the polarization-sensitive ACTPol camera \citep{niemack/etal/2010} in
2013.

ACTPol is similar to MBAC in a number of ways: it has three
separate ``optics tubes'' each of which terminates in an independent
detector
array; it uses three antireflection coated silicon lenses per tube to
feed the detectors; it is located at the Gregorian focus of the
telescope and illuminates the 2\,m diameter secondary; the passbands
are defined in part by free-space filters \citep{ade/etal/2006}; IR
radiation is blocked with free-space filters \citep{tucker/ade/2006};
and the transition-edge-sensor (TES) detectors are read out by time-division SQUID multiplexing \citep{chervenak/etal/1999}, which is implemented using the Multi-Channel Electronics
\citep[MCE,][]{battistelli/etal/2008}.  ACTPol
differs from MBAC in that the antireflection coating is achieved
through a ``metamaterial" \citep[grooving of a silicon lens
surface;][]{datta/etal/2013}; cryogenic corrugated silicon horns feed
two orthogonal polarizations \citep{britton/etal/2010,hubmayr/etal/2012};
the lower edge of the spectral passband is determined by the feed
structure's input waveguide cutoff frequency, while the upper end is defined
by free-space filters; the detector arrays
\citep{grace/etal/2014, pappas/etal/2014} are cooled to 100\,mK as opposed to 300\,mK; and a continuous dilution refrigerator cools the detector arrays instead of a $^3$He sorption refrigerator, which enables daytime observations. The receiver is 40\%
larger than MBAC in volume.

Two of the optics tubes are designed to operate at \apfreq{} and have 522 feeds (1044 TES detectors) each, 506 of
which are typically used.  In addition to the optical detectors, there are 6 dark detectors and 6
resistors for noise tests.  The third tube will use a 90/146\,GHz
dichroic detector array with broad-band cold optics \citep{datta/etal/2014} and have 1020 TES detectors coupled to 255
feeds. For the results reported here, only one \apfreq{} tube was
installed.
The second 146~GHz tube was installed in 2014 with the final tube scheduled
for early 2015.

The array sensitivity (relative to the CMB) of the 2013 observations with the
first optics tube is $\sim 19\,\uKrs$. For comparison, the MBAC
sensitivity at 148 GHz was $32\,\uKrs$. The 405 most sensitive detectors
provide 95\% of the statistical weight in the maps, and the {\it in situ} median detector
sensitivity is $340\,\uKrs$ \citep{grace/proc/2014}.
Substantial improvements in sensitivity for 2014 observations are expected due to the unusually high precipitable water vapor (PWV) in 2013,\footnote{The median nighttime PWV measured by the nearby Atacama Pathfinder Experiment weather monitor \citep[APEX,][]{gusten/etal/2006} during the ACTPol 2013 season was 1.1\,mm, while the median during the corresponding periods in 2008-2010 was 0.7\,mm.} reductions in readout noise, reduced background loading, as well as the addition of new detector arrays.

The response of the ACTPol
detectors diminishes with frequency such that it has a loading-dependent $f_{\rm 3dB}$ (the
frequency at 50\% of the peak response) of 10-200Hz.
When accounting for our scan speed and observing strategy,
$f_{\rm 3dB}=70$\,Hz corresponds to a rolloff in the angular
power spectrum of 3\% at $\ell =3000$. The individual $f_{\rm 3dB}$
frequencies of the detectors, including their dependence on
loading from the changing elevation and precipitable water vapor
(PWV), are accounted for in the analysis, although
it contributes only a few percent (0.1\% at $\ell=3\,000$; 2\% at $\ell=10\,000$)
difference in the power spectra compared to assuming the median value.

The passbands of the detectors were determined in the laboratory, before
deployment, with a Fourier Transform Spectrometer.  An effective
frequency for the CMB $\nu_{\rm CMB}=146\pm3$\,GHz was measured and we adopt this
throughout the analysis.  The error bar is systematic, and will
improve after planned measurements of the passband in the field.  The
resulting conversion between CMB and
Rayleigh-Jeans equivalent source temperatures is $\delta T_{CMB}/\delta
T_{RJ} = 1.66\pm0.04$. The Dicke bandwidth \citep{Dicke/1946} is $\Delta\nu=49$\,GHz.

\section{The observations and data reduction}
\label{sec:obs}
\subsection{Sky coverage and scan strategy}
ACTPol data are acquired by scanning the telescope
in azimuth at a variety of different
elevations. A patch is scanned as it rises in the east and then again
as it sets in the west. In this first year, we concentrated on
four ``deep fields" approximately centered on the celestial equator at
right ascensions $150^\circ$, $175^\circ$,
$355^\circ$ and $35^\circ$ which we call D1 (73\,deg$^2$), D2
(70\,deg$^2$), D5 (70\,deg$^2$) and D6 (63\,deg$^2$) respectively.
The areas refer to the deep, rectangular regions with even coverage
in the center of each patch that we use for power spectrum analysis. These
patches were chosen for their overlap with other surveys and are shown
in Figure~\ref{fig:maps}.  The separation of the patches is such that
only one is visible at any given time, and in a typical 24 hour period
all four patches were observed in sequence unless the observation would
require the telescope to point within five degrees of the Sun.
With our scan strategy, each patch is observed in
a range of different parallactic angles while scanning
horizontally. This is important for separating instrumental from
celestial polarization and is a benefit of observing from a non-polar
site.

The CMB fields are observed by scanning at $1.5^{\circ}$/s in azimuth,
turning around in 1\,s, scanning back to the original position,
turning around in 1\,s and repeating. The duration of the scan depends
on elevation; a full cycle takes 16.4\,s at an elevation of $35^\circ$
and 20.9\,s at $60^\circ$. This is done for 60 scans, or roughly 10
minutes, to form a time-ordered-data or ``TOD" packet. The elevation is sometimes
changed between 10 minute scans, at which point the
detector bias is modulated to recalibrate and check for any changes in
the time constants due to the change in sky load.

Data for the maps in Figure~\ref{fig:maps} were taken from Sept. 11,
2013 to Dec. 14, 2013.  During this time, in addition to observing the
CMB, we performed a number of systematic checks, characterized the
instrument, observed planets, and observed the Crab Nebula
(\taua{}). The net amount of time that went into the maps was 236, 178,
311, and 305 hours for D1, D2, D5, D6 respectively.
This represents 24\%, 16\%, 29\% and 31\% of the total CMB observation time.
However, we use only the lowest noise regions of the maps, which constitute around $70\%$ of
the total observing time. This results in a white noise map
sensitivity, in the sense of Figure~2 in \citet{das/etal/2014}, of 16.2,
17, 13.2, and 11.2\,$\mu$K-arcmin respectively. 
For Stokes $Q$ or $U$
sensitivities these numbers should be multiplied by $\sqrt{2}$.

We divide the data into ``day'' (11:00-24:00 UTC) and ``night''
(0:00-11:00 UTC). The nighttime data fraction for patches D1, D2, D5
and D6 is 50\%, 25\%, 76\% and 94\% respectively.
For this analysis we use only the nighttime data from D1, D5, and D6,
amounting to 63\% of the total.

\subsection{Beam, pointing, and polarization reconstruction}
\label{subsec:beam}
We have found that multiple observations of planets
\citep{hincks/etal/2010, hasselfield/etal/2013} are essential for
determining the beam profile. In 2013 Uranus was observed 120 times
and analyzed as in \citet{hasselfield/etal/2013}. With all detectors
combined, regardless of polarization, the beam is slightly elliptical
with FWHM of $1.36'$ ($1.26'$) along the major (minor) axis.
The solid angle is
$\Omega_B=194\pm6$~nsr ($2.29 \pm 0.07$\,arcmin$^2$), before
any smearing due to pointing. These results agree up to
$\ell=5000$ with a similar analysis of 20 observations of Saturn, a
much brighter source.  We did not detect any significant deviations
when the data were split by the elevation of observation or whether
the source was rising or setting. The beam profile was marginally detected in
polarization maps of Uranus made in coordinates fixed to the optical
system.  Since the observations probed the planet at a range of
parallactic orientations, this signal is interpreted as either $I$ to $P$
leakage in the optics or due to the analysis pipeline.
The leakage from $I$ to $P$ due to this effect is less than
$1.5\%$ at $\ell < 5000$.  The leakage is dominated by monopole terms
and thus is highly suppressed in CMB maps because a range of
parallactic angles is explored by the cross-linking scan strategy.
Including the effects of cross-linking in the analysis of Uranus, we
find the polarized fraction of the 146~GHz emission from Uranus to be less
than 0.8\% at 95\% confidence.

A simple telescope pointing model is constrained using observations of
planets at night.  This model allows the pointing to be
reconstructed with an rms error of $14''$.  The impact of pointing
variance on the full season CMB maps is handled as
in \citet{hasselfield/etal/2013}, leading to effective beams in the
CMB maps with solid angle $\Omega_B=224 \pm 20$\,nsr for D1,
$\Omega_B=234\pm14$\,nsr for D5, and $\Omega_B=224\pm12$\,nsr for D6.
The uncertainties include both beam and pointing contributions.

Average pointing error in the full season maps is assessed by
comparing point source positions to the FIRST catalog
\citep{becker/etal/1995,white/etal/1997}.  The absolute pointing
error rms is found to be $5''$ in the nighttime maps, with no significant
deviations when the data are split by elevation or time of
observation.  A $7''$ offset in the absolute pointing is
uncorrected in this analysis.

Detectors co-located in the focal plane may point at slightly
different positions on the sky, an effect seen by the
\citet{bicep2b/2014} at the $1'$ level.  Because our detector
pointing offsets are individually measured, and the map making
procedure does not difference detectors directly (\S\ref{sec:ds}),
this is not a primary concern for us but it is an important check of
the instrument.  From the analysis of planet observations, we find
that the optical axes of a typical detector pair differ by less than $3''$.

During the day, the heat from the Sun distorts the telescope
structure. The distortion pattern is repeatable, although we currently cut data
between 17:00 and 20:00 UTC (13:00 and 16:00 local time) when the distortion
is greatest. The distortion leads to two effects: the first is a pointing offset and the second is a repeatable deformation of the reflector, which changes the shape of the beam. Both of these effects lead to a roll-off in $\ell$ that resembles a low pass filter and can be treated as a beam effect in the likelihood.
We do not include the daytime data in our cosmological parameter
analysis, as models for the daytime beam response are still in
development.
Nevertheless, our preliminary treatment of the
daytime beam produces power spectra that are consistent with nighttime
spectra (Figure~\ref{fig:null})  and pass null tests (see \S\ref{sec:nulls}).

Each detector is sensitive to a single linear polarization direction,
and the relative angles of the detectors within each array are set by lithography during
fabrication.  The optical detector orientations are calculated by raytracing through the full
optical system to determine the projection of each detector within an array on the sky.

CMB temperature fluctuation measurements from \planck{} and \map{} are
used to calibrate the ACTPol T, Q, and U maps with a common rescaling
factor, as described in \S\ref{sec:pspec}.  However, signal in the Q and U
maps will be attenuated by an additional factor $\alpha_\text{P}$,
which is different from unity due primarily to errors in the assumed detector
polarization angles.  For the present analysis we take
$\alpha_\text{P} = 0.95 \pm 0.05$; this is discussed further
in \S\ref{sec:taua}.

Any mean offset between the assumed and actual detector polarization angles
must be understood in order to properly decompose polarized
intensity into E and B components.
The polarized CMB may be used to assess the offset angle under the
assumption that the intrinsic correlation between the E and B
signals is zero.  Systematic optical effects associated with
polarization, including parallactic rotations, cause a leakage from
E to B modes and induce spurious signal in EB and TB
correlations (e.g., \citealt{shimon/etal/2008}).  The most likely
instrument polarization reference angle may thus be determined by
minimizing the inferred EB signal with respect to offset angle
(e.g., \citealt{keating/etal/2013}).
Under these assumptions, the ACTPol E and B spectra from $500 < \ell <
2000$ constrain the instrumental polarization offset angle to be
$\delta\gamma_p = -0.2^\circ \pm 0.5^\circ$.  This result is referred
to as the EB-nulling offset angle.  Since this angle is small and
consistent with zero, we do not correct the spectra for this effect in
the present analysis.  The agreement with zero suggests that the
optical modeling procedure is free of systematic errors at the
$0.5^\circ$ level or better.

Naturally, estimating the polarization offset angle by assuming E and B
to be uncorrelated eliminates sensitivity to models, such as
isotropic cosmic birefringence, where the distinguishing characteristic
is a constant EB cross-correlation.  
As an alternative calibration approach, measurements of the polarized signal from
a bright astrophysical source may be compared to values from the
literature. The Crab Nebula is a convenient source for this purpose.

\subsection{The Crab Nebula}
\label{sec:taua}

\begin{figure}
\includegraphics[width=\columnwidth]{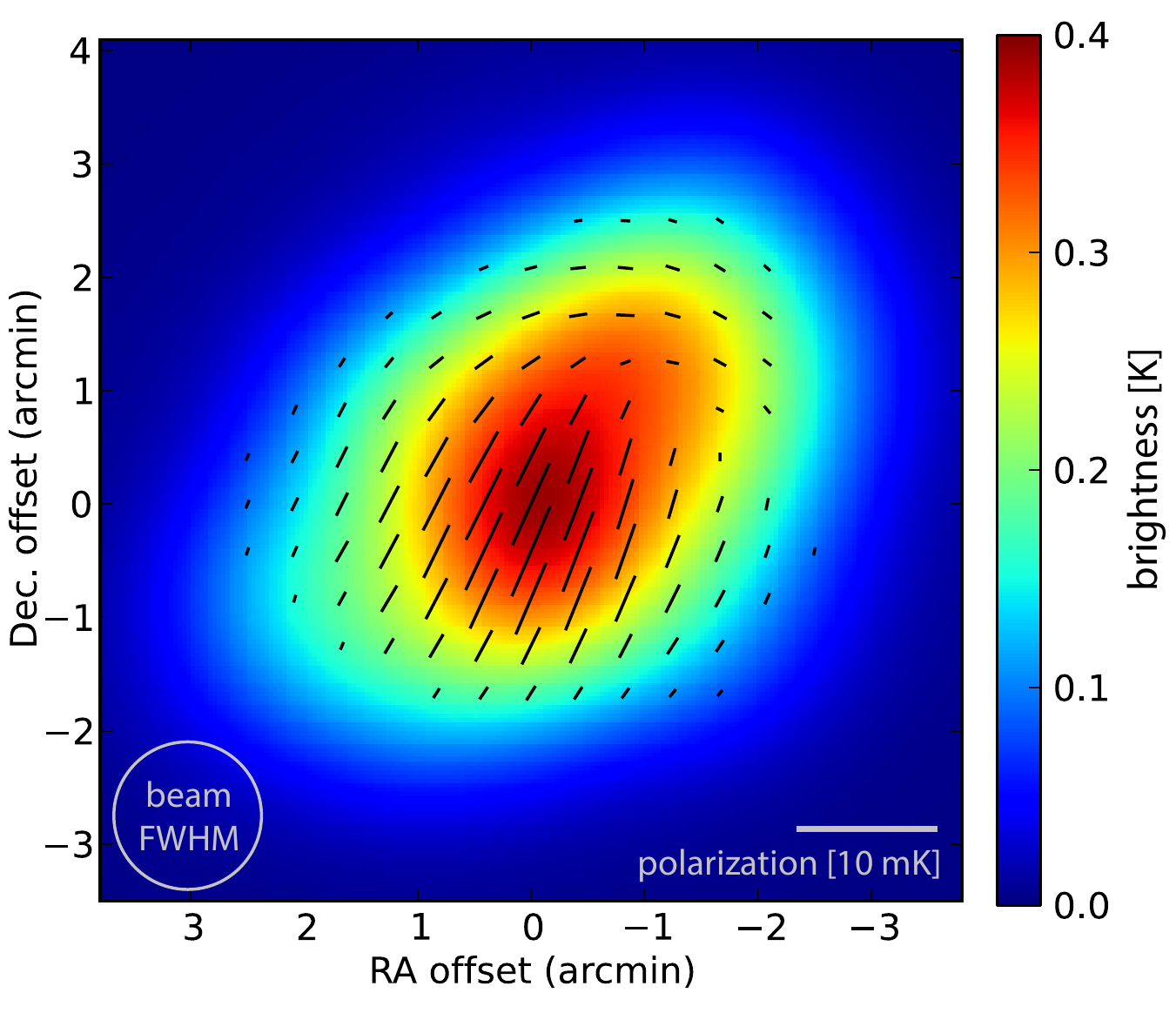}
\caption{ACTPol intensity and polarization map of the Crab Nebula
(\taua{}).  The brightness scale is in differential temperature units,
  relative to the CMB at \apfreq{} (for Rayleigh-Jeans brightness
  temperature, divide by 1.66).  Black bars show
  polarization direction, with length proportional to polarized
  intensity ($(Q^2 + U^2)^{1/2}$), which reaches a maximum of 6.8\,mK.
  The origin of the coordinate system corresponds to the Crab Pulsar
  position.  Note the variation in polarization angle and fraction
  across the source, which complicates the use of the source as a
  precision calibrator.  Polarization features are in good
  qualitative agreement with the higher resolution, 89.2 GHz maps of
  \citet{aumont/etal/2010}; a numerical comparison is discussed in the
  text.  This map has been resampled to $2''$ pixels from its original
  $30''$ resolution but has not otherwise been smoothed.}
\label{fig:taua}
\end{figure}

The Crab Nebula, or \taua{}, is an extended polarized
supernova remnant whose emission below \mbox{$\approx1$\,THz} is dominated
by synchrotron radiation \citep[see, e.g.,][]{hester/2008}.
Its brightness and compactness provide a convenient
polarization reference for millimeter-wavelength observatories.  From
\map\ measurements at 93 GHz, \taua{} is polarized in the direction
$\gamma_p = 148.9^\circ \pm 1.8^\circ$
in equatorial coordinates \citep{weiland/etal/2011}. These
agree with the IRAM observations at 89.2 GHz
which give the direction as $\gamma_p= 149.9^\circ\pm0.2^\circ$
when smoothed to a $5'$ Gaussian beam \citep{aumont/etal/2010}.

\taua{} was observed roughly every second day during the 2013 observing
season.  A co-added map from a subset of these observations is shown
in Figure~\ref{fig:taua}.  The results presented here are a
preliminary analysis with the goal of demonstrating the level to which
the polarization sensitivity of the instrument is understood.

Because of their $27''$ resolution and higher precision, we take the
IRAM results as our reference.  We downgrade the resolution of the ACTPol maps to
match a $5'$ Gaussian beam and compare to the \citet{aumont/etal/2010}
values for this case.\footnote{We are collaborating with the \planck\
  team to develop \taua{} as a standard but this present analysis used
  only public IRAM results.}  The
difference corresponds to an instrument polarization offset angle of
$\delta\gamma_p = -1.2^\circ \pm 0.2^\circ$ at the pulsar position in the
smoothed maps, where the uncertainty is statistical and is assessed by
comparing three independent sets of ACTPol observations.

The analysis of \taua{} was performed while keeping the EB result
``blinded,'' and provides an independent probe of the
instrumental polarization offset angle.
With the EB result unblinded, the difference between the IRAM-based
and EB-nulling polarization offset angles is $1.0^\circ \pm 0.5^\circ$.
This difference is consistent with the $1.1^\circ \pm 0.5^\circ$ offset
found by \citet{pbear-eebb/2014} at 150 GHz, and increases the
evidence of a $\approx 1^\circ$ difference in \taua{} polarization
between 90 and 150\,GHz.  (The uncertainty stated here for
the \pbear{} result does not include a contribution from IRAM overall
polarization uncertainty.)  

Studies of the total flux density of the nebula have shown that the
spectrum is well described by a single synchrotron
component \citep{macias-perez/etal/2010}, and predict negligible
variations in polarization fraction and angle at 150\,GHz.  However,
high resolution studies of the source at 150\,GHz and 1.4\,GHz
demonstrate non-negligible variations of the spectral index over the
surface of the source \citep{arendt/etal/2011}.  Since the polarization
angle also varies over the source, the total polarized flux and mean polarization angle may have a non-trivial behavior as a function of frequency.

Our understanding of the polarization efficiency is currently limited
by uncertainty in individual detector polarization angles.  Comparison
of individual detector timestreams to the ACTPol \taua{} maps provides
limits on these angle errors.  At the present time we can only
conclude that the polarization efficiency $\alpha_\text{P}$ is at
least 0.9, corresponding to an rms uncertainty in the polarization
angles of $10^\circ$.  While this is somewhat larger than the expected
deviations, for the present analysis we consider the full range $0.9
< \alpha_\text{P} < 1.0$.  In the cosmological likelihood analysis,
the efficiency is given a uniform prior over this interval.  When
stating polarization fractions below, we simply take $\alpha_\text{P}
= 0.95 \pm 0.05$ and treat the error as Gaussian.

For the purposes of stating our measurements of the \taua{}
polarization signature, we apply the EB-nulling instrumental
polarization offset.  These results apply at \apfreq{}, for an
instrument with a $5'$ Gaussian beam.  At the pulsar position, the
polarization fraction is $(9.2 \pm 0.5)\%$ and the polarization
angle is $150.9^\circ \pm 0.5^\circ$ East of North.  At the peak of the smoothed
intensity (which lies $22''$ northwest of the pulsar position), the
polarization fraction is $(8.7 \pm 0.4)\%$ at an angle of $150.7^\circ \pm
0.6^\circ$.

\subsection{Data selection and pre-processing}
\label{sec:ds}

The data selection closely follows the path laid out in
\citet{dunner/etal/2013}.  The main difference in the TOD processing is the procedure for calibrating each detector, as a larger fraction of the
detectors are operating near the saturation level.
A set of reliable detectors
is used to determine an absolute calibration level that is stable over
variations in detector temperature and loading.  A low-order polynomial is
removed from each time stream to limit the impact of low frequency drifts.
Then a per-TOD flat-fielding is performed on all the detectors, using the
common-mode signal from the atmosphere.  At this point the properties of the
detector time streams are characterized and screened.  The output of this step
in the pipeline, which we call the cuts package, is a list of science grade
detectors with $f_{3\textrm{dB}} > 20 \textrm{Hz}$, well-behaved noise, and a
common relative calibration.  In addition, we reject data when the PWV$ >
3$\,mm.  Note that the polarization orientation of a detector does not enter
into the data selection or flat-fielding. The time stream processing (such as
polynomial removal) used to determine the cuts and calibration are not the same
as those applied to the data during mapmaking.

\begin{figure*}[htb]
	\centering
	\includegraphics[width=\textwidth]{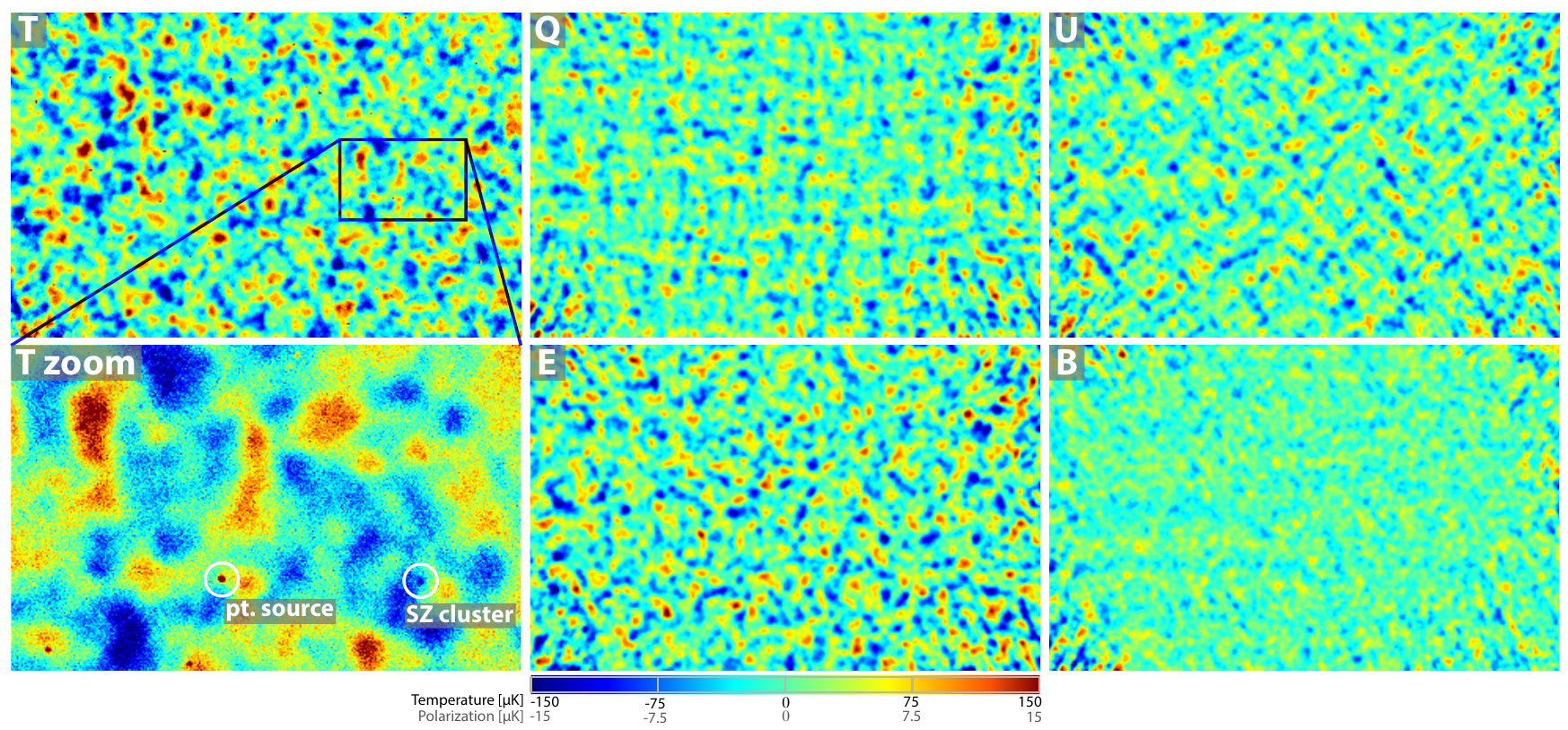}
	\caption{Example maps from the region $29.65^\circ < \textrm{RA} < 40.49^\circ$ (horizontal),
		$-7.60^\circ < \textrm{Dec} < -0.68^\circ$ (vertical), in the center of patch D6.
		Panels 1,2,3,5,6 (left to right, top to bottom) show T, Q, U, E and B
		respectively. Panel 4 is a zoom on
		a $2.79^\circ \times 1.73^\circ$ subregion of the T map, showing the full
		map resolution. The maps have been bandpass
		filtered to maximize signal-to-noise ($240<\ell$ for temperature, $260<\ell<1370$
		for polarization). The visible patterns in the Q and U maps are consistent with
		a sky dominated by E-mode polarization, as can be seen in the derived E and B maps.
		The B map is consistent with noise except for a faint
		$m=0$ (constant declination) ground residual (see \S\ref{sub:pickup}).
		We do not use $m=0$ modes in the power spectrum estimation.
		See Figure~\ref{fig:diff6} for an illustration of the
		noise properties in these filtered maps.
		The circled galaxy cluster candidate, ACT-CL
		J0205.2-0439, is within $2'$ of a CFHTLS cluster
		candidate with photometric redshift $z =
		1.1$ \citep{durret/etal/2011} and three concordant
		galaxies with spectroscopic $z = 0.97$ found in the
		VIMOS Public Extragalactic
		Survey \citep{garilli/etal/2014}.  The circled point
		source may be associated with FBQS J0209-0438, a
		quasar at $z = 1.128$ \citep{veron-cetty/veron/2006}.}
	\label{fig:deep6}
\end{figure*}

\begin{figure*}
	\centering
	\setlength{\tabcolsep}{1.5pt}
	\includegraphics[width=\textwidth]{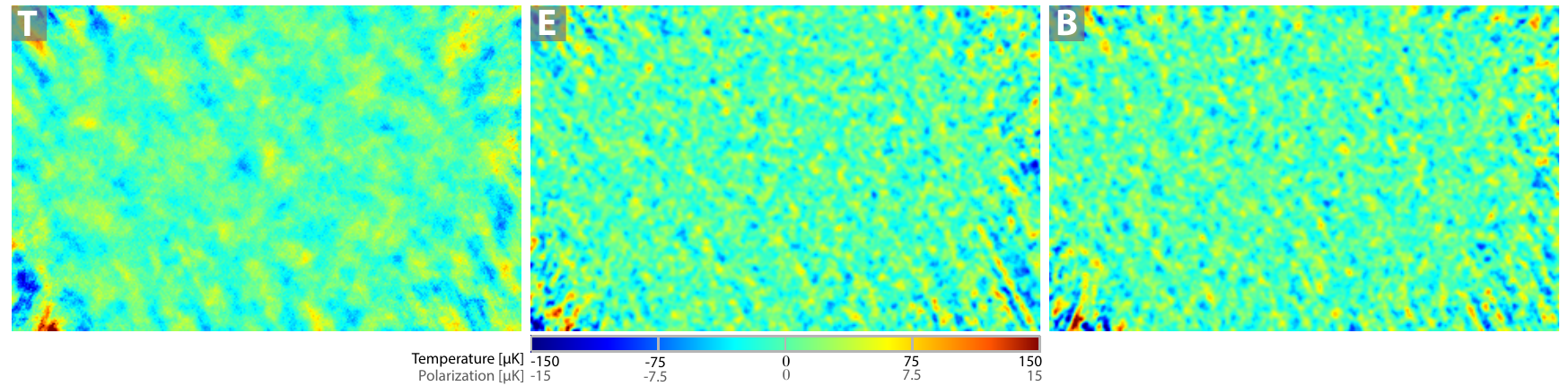}
	\caption{Difference maps (odd vs. even pairs of nights) for the same region
		as Figure~\ref{fig:deep6}, illustrating the noise properties of the
		map. Correlated noise is visible as diagonal stripes
		aligned with our dominant
		scanning directions (diagonally in these coordinates). These are the map-space equivalent of the correlated
		noise in the time-ordered data. Noise correlations are taken into account
		in the noise model in the power spectrum estimation.}
	\label{fig:diff6}
\end{figure*}

\subsection{Map making}
After applying the cuts, the time-ordered data are projected on the sky
by solving the maximum-likelihood map making equation for a vector of
map pixels $\mathbf{m}$,
\begin{align}
	A^TN^{-1}A\mathbf{m} = A^TN^{-1}\mathbf{d},
\end{align}
via the preconditioned\footnote{We currently use a simple binned preconditioner,
	with a 3x3 matrix at each $(I,Q,U)$ map pixel that inverts the
	(polarized) hit-count map.
	We are investigating other approaches such as a stationary correlation
preconditioner \citep{naess/louis/2013}.} Conjugate Gradients algorithm. Here $\mathbf{d}$
is the set
of time-ordered data, $N$ is the noise covariance of $\mathbf{d}$ and $A$ is the generalized
pointing matrix that projects from map domain to time domain. This follows the method
used in \citet{dunner/etal/2013}, but extends it to polarization.
Polarization is handled by including each detector's response to the I, Q and U
Stokes parameters
in the pointing matrix, so the analysis does not depend on explicit detector
pair differencing.
This approach, coupled with the parallactic angle coverage of our scan strategy,
naturally suppresses monopole and dipole polarization leakages.
The different noise properties for temperature and polarization
are represented as detector correlations in the noise covariance matrix,
which we model as stationary in $10$ minute chunks and measure from the data.\footnote{
In temperature, the atmosphere is our largest noise term for $\ell\lesssim3\,000$, but
atmospheric noise is almost absent in polarization.
}
To avoid bias from applying the noise model to the same data it is measured from,
we make a second pass where the estimated sky map is subtracted from the time-ordered
data before the noise model is reestimated.

In principle, maximum-likelihood map making results in unbiased,
minimum-variance sky maps.
But there are a few caveats.
With the Conjugate Gradients technique,
the number of steps needed to solve for each eigenmode depends on its eigenvalue, and
some degenerate or almost degenerate modes never converge. The nature of the
degenerate modes depends on the patch size and scanning pattern, but in the
present analysis they correspond to the low signal-to-noise modes at
multipoles $l \lesssim 50$.
Additionally, our current treatment of ground pickup results in a bias on
these large scales (see \S\ref{sub:pickup}).

With our current data set, each 70 deg$^2$ map is a reduction of $\sim 10^{11}$
samples into $\sim 10^6$ pixels, making this the most computationally intensive
step in the analysis.
Nevertheless, due to the short observation time so far, the costs are
still relatively modest compared to the original ACT analysis.
Figure~\ref{fig:deep6} shows an example map from patch D6.
 For
display purposes, a bandpass filter has been applied to maximize signal-to-noise.
Example difference maps (odd vs.\ even pairs of nights)
for the same region are shown in Figure~\ref{fig:diff6}.

\subsection{Ground, lunar, and solar pickup}
\label{sub:pickup}
ACT has two levels of ground screens. One screen is fixed to the telescope and
scans with it. This entire system sits inside a second 13~m high fixed ground
screen. Nevertheless, we still detect a spurious signal which we interpret as
ground pickup. When the telescope points to the northeast between azimuths
$\sim 25^\circ-85^\circ$ we observe a spurious signal with a $\sim 30^\circ$
period in azimuth, with little elevation dependence, and a peak-to-peak
amplitude of $\sim 200 \mu$K in Q and U. This is
consistent with signal from the nearby mountain Cerro Toco being diffracted
over the top of the ground screen's $30^\circ$ wide panels. While this signal
is washed out when projected on the sky, it would still be a contaminant of
$\pm 20 \mu$K or more in our polarization maps if ignored.

The ground signal can be disentangled from the sky because it
is constant in azimuth during a scan and does not rotate with parallactic
angle like the sky. The exceptions are modes on the sky that depend only on
declination, not right ascension (such as the spherical harmonics with
$m=0$). These show up as pure functions of azimuth during a constant
elevation scan, and are degenerate with the ground even when observing
at multiple azimuths and elevations.
The remaining modes could, in principle, be disentangled, but
in the current analysis we remove both these and the
degenerate modes by applying an azimuth filter to the time-ordered data
and excluding Fourier modes with $|\ell_y|<50$ from the power spectrum
estimation.\footnote{Excluding $|\ell_y| < 50$ removes an approximately $4\mu$K$/^\circ$
residual ground gradient from the azimuth-filtered maps.}

While the filters are effective at suppressing the ground pickup, they also remove some bona fide sky signal,
making our maps and power spectra slightly biased.  The effects of the
filtering are assessed by passing simulated maps of the polarized CMB through
the filtering procedure, and comparing the power spectra of the input
and output maps.  The main effect of the filter is to suppress,
slightly, the signal in temperature (polarization) on large angular
scales, with a transfer function that decreases from 0.995 (0.99) at $\ell =
500$ to 0.95 (0.9) at $\ell = 200$.  Leakage from E to B is also seen, but
at a level that is negligible for this analysis.  Our simulations show
that with a more sophisticated treatment we can expect a significantly
reduced impact from ground signals in future ACTPol results.

We investigate the possibility of contamination from sidelobes overlapping
the Sun or Moon by making maps in coordinates centered on these objects.
We identify two sidelobes this way, one around $20^\circ$ away from the
boresight, and another one $120^\circ$ away. These have an amplitude of
around $200\mu$K in polarization for the Sun, but are not detected for the
Moon. Based on this, we cut all scans that hit these regions in
Sun-relative coordinates. As a precautionary measure, we also cut all scans
within $10^\circ$ of the Moon, which results in a negligible loss of data.

\begin{figure}
	\includegraphics[width=\columnwidth]{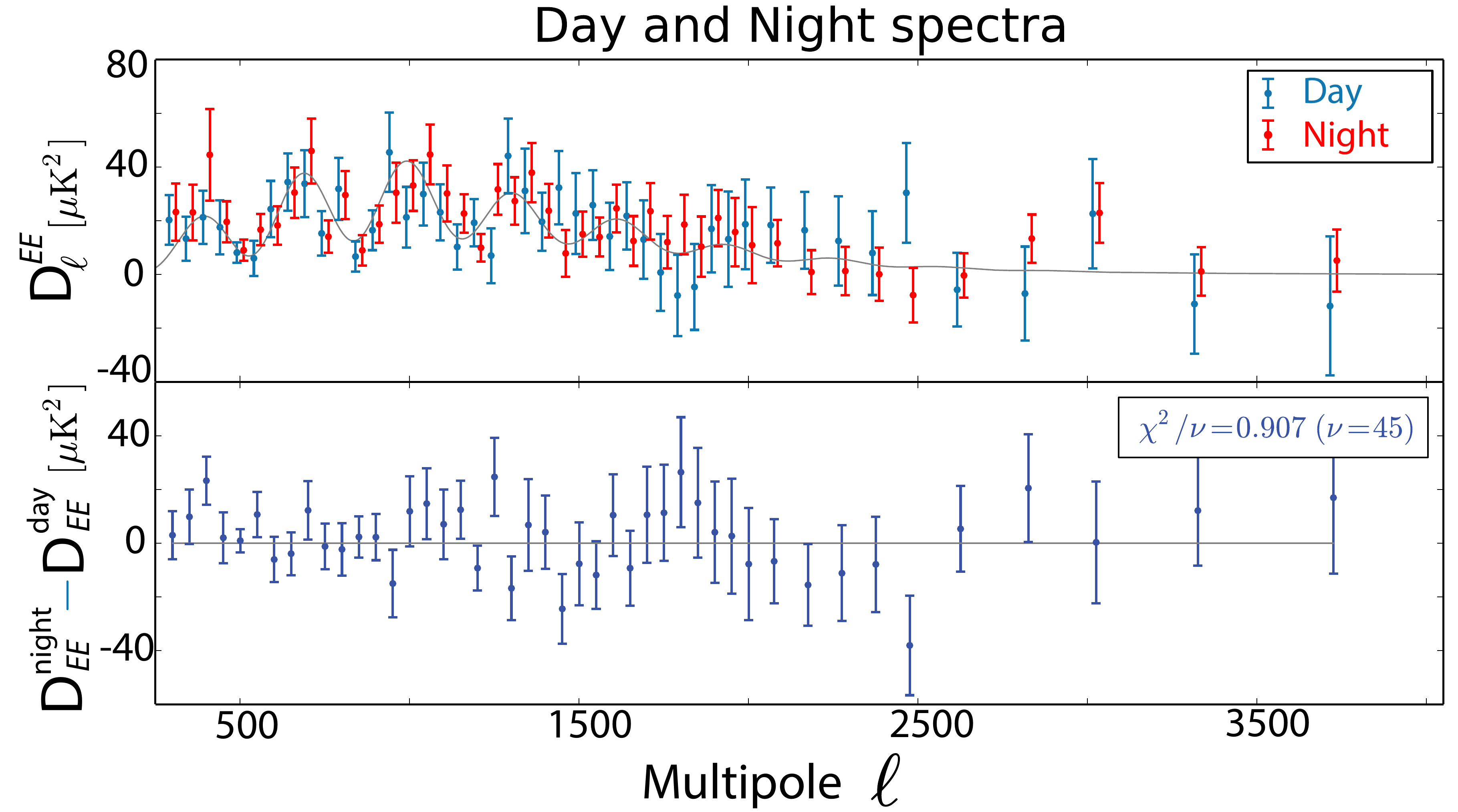}
	\caption{The D1 day and nighttime EE spectra (top), and their difference (bottom); they are consistent.}
	\label{fig:null}
\end{figure}

\section{The power spectra and interpretation}
\label{sec:pspec}

We compute the temperature and polarization power spectra of the maps following the methods described in \citet{louis/etal/2013}, which builds on those used in \citet{das/etal/2011,das/etal/2014}.
The code was checked extensively with simulations and is able to extract an E/B mode power spectrum signal to within 0.1$\sigma$ in maps with $5~\mu$K noise in the presence of uneven weights across the maps, irregular boundaries, and cutouts for sources.

In this analysis we use $\ell^{\rm TT}_{\rm min}=500, \ell^{\rm TE, EE}_{\rm
min}=250$, and $\ell^{\rm TT, TE, EE}_{\rm max}=9\,000$.\footnote{We use a larger
value of $\ell_{min}$ for TT because atmosphere $1/f$ modes impact the T maps
much more than the polarization maps.}  In two-dimensional Fourier
space we mask a vertical strip with $|\ell_x|<90$ \citep[as
in][]{das/etal/2011,das/etal/2014}, and a horizontal strip with $|\ell_y|<50$,
as described in \S\ref{sub:pickup}. The maps are calibrated to the {\it
Planck} 143~GHz temperature map \citep{planck_mission/2013} following the
method in \citet{louis/etal/2014}, resulting in a 2\% uncertainty. The overall
calibration measured by \planck\ and \map\ disagree \citep{planck_params/2013},
so we then multiply all maps by a factor of 1.012 to correspond to the more
mature \map\ calibration.  As described in \S\ref{sec:taua}, the polarization
maps are assumed to have an additional 5\% calibration uncertainty.

We test the parameter extraction and power spectrum pipeline as follows. We
generate a simulated\footnote{To include the effects of our flat-sky approximation
	in our simulations, each simulated map is generated on the full, curved sky and
then projected to our native cylindrical equal-area maps.} sky in temperature and polarization with the \map{}+ACT
parameter set \citet{calabrese/etal/2013}\footnote{Throughout this work, the
WMAP+ACT parameters differ slightly from the results of \citet{calabrese/etal/2013}
as a result of incorporating the finalized ACT beam window functions
\citep{hasselfield/etal/2013}.} and with a realistic level of unresolved point source power,
extract the portion corresponding to the ACTPol coverage for each patch, and
add a realization of the full inhomogeneous noise model as measured. We then
take the power spectra of the maps, and from the power spectra derive
cosmological parameters, marginalizing over foreground parameters. We run this
process 100 times
and recover the input cosmology to within $0.1\sigma$. We use the simulations to construct the covariance matrix for the data.

In this analysis we do not account for cosmic aberration \citep{jeong/etal/2014}, super sample lensing \citep{manzotti/hu/benoit/2014}, or the effect of the flat-sky approximation. We have simulated these sub-dominant effects and find that they have a negligible impact on derived \LCDM\ cosmological parameters; however, we are extending our pipeline to account for them in future ACTPol analysis.

\begin{table}
\caption{\small Null tests}
\begin{center}
\begin{tabular}{ll  r  r r }
\hline
\hline
Test & Patch & Spectrum &$\chi^{2}$/dof &P.T.E  \\
\hline
Detector   & D5    &TT & 0.92 & 0.65  \\
           &       &EE & 1.51 & 0.01  \\
           &       &TE & 0.65 & 0.98  \\
\\
           & D6    &TT & 1.39 & 0.03  \\
           &       &EE & 0.91 & 0.66  \\
           &       &TE & 1.02 & 0.43  \\
\hline
Turnaround\footnotemark[1]
           & D5    &TT & 0.76 & 0.91  \\
           &       &EE & 0.71 & 0.95  \\
           &       &TE & 0.86 & 0.76  \\
\\
           & D6    &TT & 0.92 & 0.65  \\
           &       &EE & 1.18 & 0.17  \\
           &       &TE & 0.76 & 0.91  \\
\hline
Splits\footnotemark[1]
           & D5    &TT & 0.67 & 0.97  \\
           &       &EE & 0.72 & 0.94  \\
           &       &TE & 0.55 & 0.997 \\
\\
           & D6    &TT & 0.94 & 0.60  \\
           &       &EE & 0.77 & 0.89  \\
           &       &TE & 0.77 & 0.89  \\
\\
           & D1    &TT & 1.60 & 0.003 \\
           &       &EE & 0.77 & 0.89  \\
           &       &TE & 1.14 & 0.23  \\
\hline
Patches    & D1-D5 &TT & 0.89 & 0.70  \\
           &       &EE & 0.89 & 0.70  \\
           &       &TE & 1.24 & 0.11  \\
\\
           & D1-D6 &TT & 0.67 & 0.97  \\
           &       &EE & 0.66 & 0.97  \\
           &       &TE & 1.26 & 0.09  \\
\\
           & D5-D6 &TT & 0.94 & 0.60  \\
           &       &EE & 0.94 & 0.60  \\
           &       &TE & 0.96 &0.56   \\
\hline
Day-Night  & D1$^N$-D1$^D$ & EE & 0.91 & 0.64 \\
\\
\hline
\end{tabular}
\end{center}
\footnotetext[1]{Reported for $(s^{t}_{0}-s^{nt}_{1}) \times (s^{t}_{2}-s^{nt}_{3})$ for the turnaround nulls, and $(s_{0}-s_{1}) \times (s_{2}-s_{3})$ for the split nulls. The other permutations are reported in Table~\ref{tab:nulls_full}.}
\label{tab:nulls}
\end{table}

\begin{table}
\caption{Split and turnaround nulls }
\begin{tabular}{ccrrrrr}
\hline
\hline
Patch & Combination & Spec.&$\chi^{2}$/dof &P.T.E &$\chi^{2}$/dof &P.T.E\\
 &&& Splits & & Turn.\footnotemark[1]  \\
\hline
D5              & $(s_{0}-s_{2}) \  \times $  &TT & 0.65 &  0.98 & 0.64 &    0.98   \\
              &   $ (s_{1}-s_{3})$  & EE & 0.88 &   0.72 & 0.86 &   0.77  \\
              &    & TE & 0.77 &    0.90    & 0.76 &   0.90 \\
&&&&&&              \\
 	     & $(s_{0}-s_{3}) \  \times $ &TT & 1.08 &   0.31  & 1.09 &   0.31 \\
              &   $(s_{1}-s_{2})$   & EE & 0.92 &  0.64 & 0.95 &   0.58   \\
              &    	  & TE & 1.21 &   0.14 & 1.34 &   0.05 \\
\hline 	
D6              & $(s_{0}-s_{2})\  \times $  &TT & 0.93 &   0.61  & 1.17 &    0.19  \\
              &    $(s_{1}-s_{3})$	   & EE & 0.60 &   0.99  & 0.75 &   0.92   \\
              &      & TE & 0.72 &  0.94    & 0.79 &   0.87    \\
&&&&&&              \\
 	     & $(s_{0}-s_{3}) \  \times $ &TT & 0.96 &   0.55 & 1.06 &   0.36  \\
              &    $ (s_{1}-s_{2})$	 & EE & 0.74 & 0.92   & 1.06 &   0.35   \\
              &    	    & TE & 0.91 &  0.67  & 1.01 &   0.46   \\
\hline 	
        D1      & $(s_{0}-s_{2}) \  \times$  &TT & 0.68 &   0.97   \\
              &   $ (s_{1}-s_{3})$  & EE & 0.89 &   0.71    \\
              &     & TE & 1.16 &  0.20      \\
 &&&&&&             \\
 	     & $(s_{0}-s_{3}) \ \times $ &TT & 0.92 &   0.65   \\
              &  $(s_{1}-s_{2})$  & EE & 0.69 & 0.96    \\
              &      & TE & 1.30 &  0.06    \\
\hline
\end{tabular}
\footnotetext[1]{For turnarounds the second split-map in each difference has the turnaround removed, e.g., the first row has $(s^{t}_{0}-s^{nt}_{2}) \times (s^{t}_{1}-s^{nt}_{3})$.}
\label{tab:nulls_full}
\end{table}

\subsection{Null tests}
\label{sec:nulls}
We perform a set of null tests similar to those done for the ACT temperature
analysis \citep{das/etal/2014}: compare the data with and without telescope
turnaround periods incorporated,\footnote{Labeled $s^t$ and $s^{nt}$ respectively.}
compare the results from different detector
sets, and compare the maps made from the four different time-splits.\footnote{$s_i$ for
$i$ from 0 to 3, such that $s_i = \{j:i=j \mod 4 \}$, where $\{j\}$ is the
set of all observations in chronological order, i.e. a 4-way equivalent of an odd-even split.}

For the detector null, we split the array in half by those most likely to have a different calibration and polarization response, and make two split maps for each subset. We form the difference map between detector sets for each split
and then compute their cross-spectrum. For the turnaround null we test for effects generated by the acceleration of the telescope. As in \citet{das/etal/2014} we make four split maps with the turnaround data removed, cutting 12-13\% of the data, and form a set of difference maps between splits with and without turnarounds. We apply these tests to the D5 and D6 patches, which have the deepest coverage.

We have also compared the daytime and nighttime spectra for the D1 patch, and
find them to be consistent (see Figure~\ref{fig:null}). The null tests are
summarized in Table ~\ref{tab:nulls}, and Table \ref{tab:nulls_full} shows the
other permutations of the null spectra for the turnaround excision and split
nulls.

An assessment of the consistency of the $\chi^2$ and probability-to-exceed (PTE)
statistics is complicated
by the fact that the many tests probe the same noise realization and are thus
correlated. For independent measurements, the distribution of PTE values
should be consistent with a uniform distribution. Our distribution of PTE
values is somewhat skewed towards values greater than 0.5. However, there is no
systematic failure of the $\chi^2$ test in these results, and the most extreme
values of 0.003 and 0.997 are not statistically surprising for a sample of this
size.
The lowest PTE is for a D1 TT split null
(0.003), and the highest for a D5 TE split null (0.997), but the other
permutations of these null spectra, shown in Table \ref{tab:nulls_full}, do not
show outlier behavior.

\begin{figure}
	\centering
	\includegraphics[width=\columnwidth]{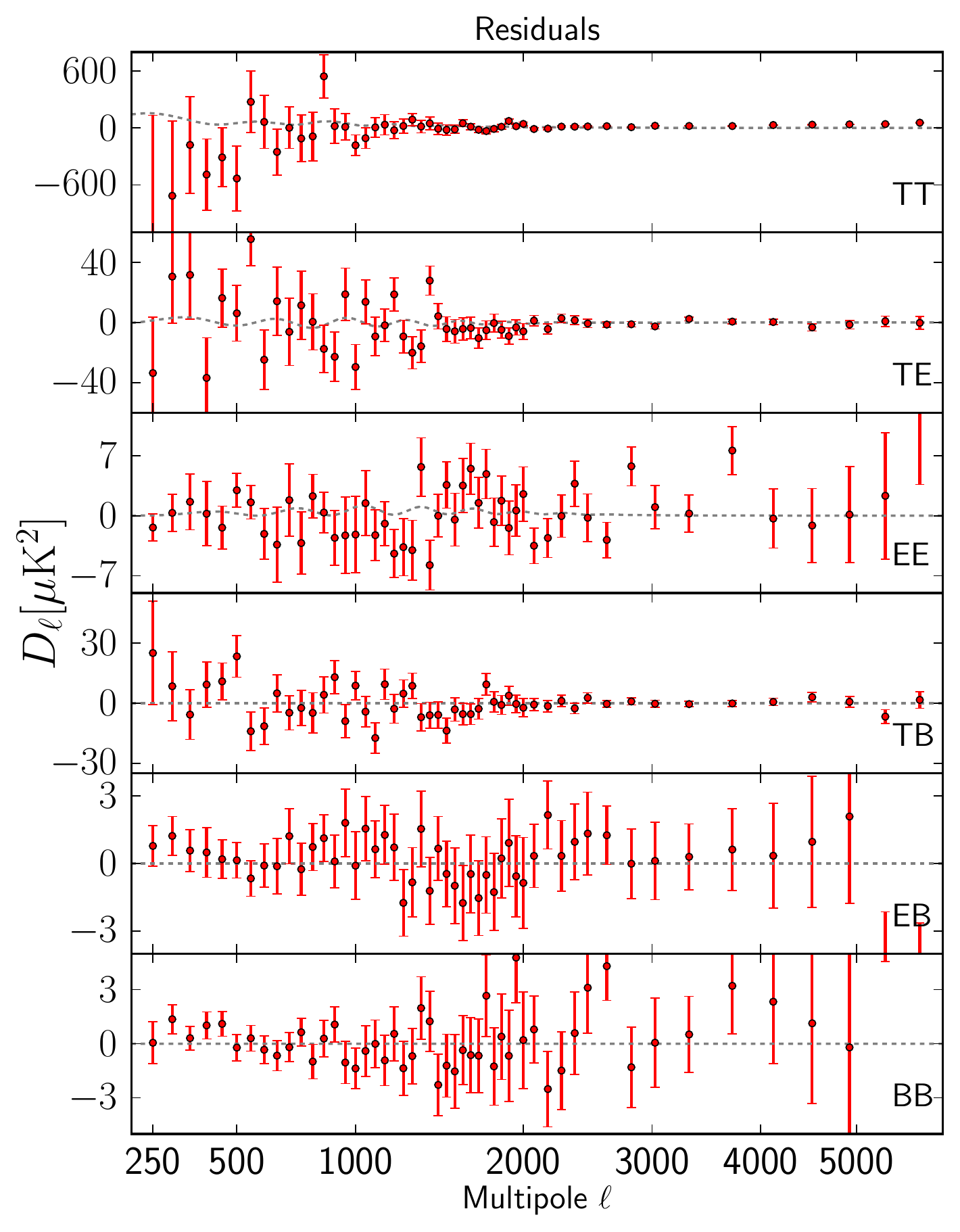}
\caption{
Residuals of measured power spectra relative to the \map{}+ACT best-fit
model (for which TB, EB, and BB are assumed to be zero).  Dashed curve
shows the small difference between the \map{}+ACT and Planck+WP+highL
best-fit models. The $x$-axis is scaled as $\ell^{0.5}$.
}
\label{fig:resids}
\end{figure}

The different patches have different coverage, are observed at different times, have different (but low) levels of potential Galactic contamination, and are observed differently relative to the local environment. 
For the analysis of many systematic effects, they are effectively independent measurements, so the spectra can be compared as an additional test.
Figure~\ref{fig:resids} shows 
the combined power spectra from the three patches. For the spectra where a signal is detected (TT, TE, EE) we have subtracted the \map{}+ACT best-fit model \citep[reproduced from]{calabrese/etal/2013} and show the residuals. For reference we show the small difference between the \map{}+ACT model and the Planck+WP+highL model for TT, TE and EE.  For TB, EB, and BB we just show the data. The measured BB signal is consistent with zero as expected with the current ACTPol sensitivity.
We find that the spectra are consistent among patches, with the $\chi^2$ of their differences given in Table~\ref{tab:nulls} for TT, TE, and EE.

\begin{figure}
	\includegraphics[width=\columnwidth]{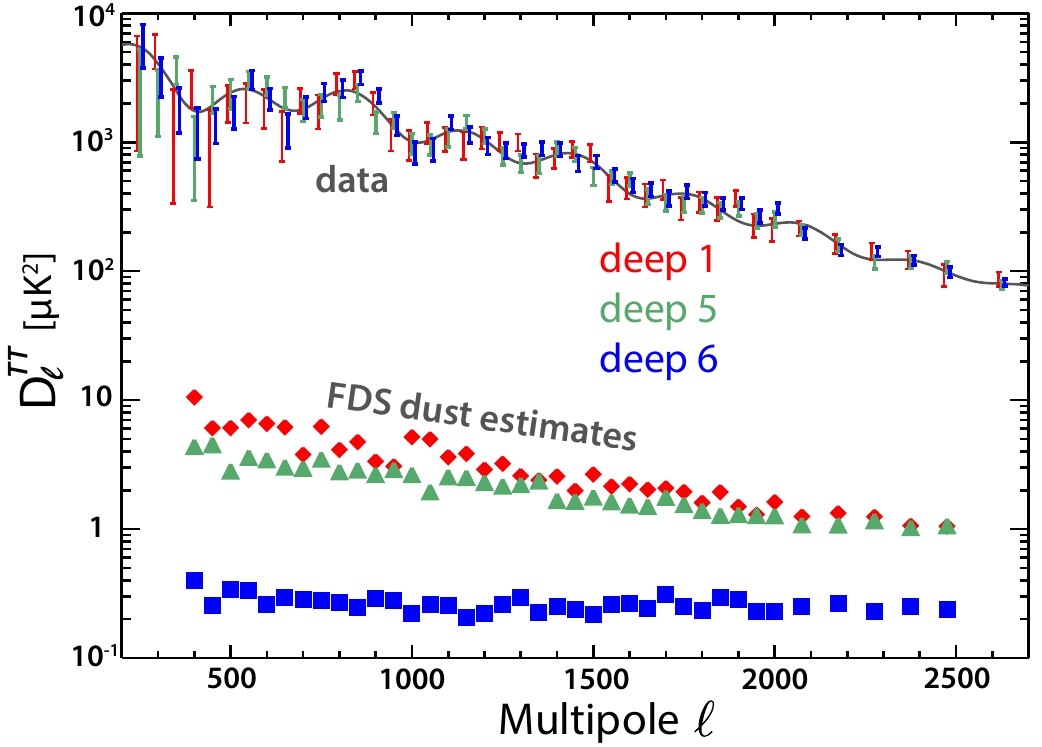}
	\caption{The expected temperature power spectrum of thermal dust in the ACTPol patches, estimated using the \citet[][FDS]{finkbeiner/davis/schlegel/1999} template, plotted below the ACTPol temperature power spectra.  The dust amplitude is $\leq 2$ $\mu$K$^2$ at $\ell = 2000$ ($\sim 1$\% of the TT spectrum amplitude, or 10\% in the maps).
	We show that the FDS template is a good tracer of the sub-dominant dust component by correlating it with the ACTPol maps, finding a cross-correlation consistent with unity to within 1$\sigma$.}
\label{fig:dust}
\end{figure}

\begin{figure*}
	\centering
	\includegraphics[width=0.82\textwidth]{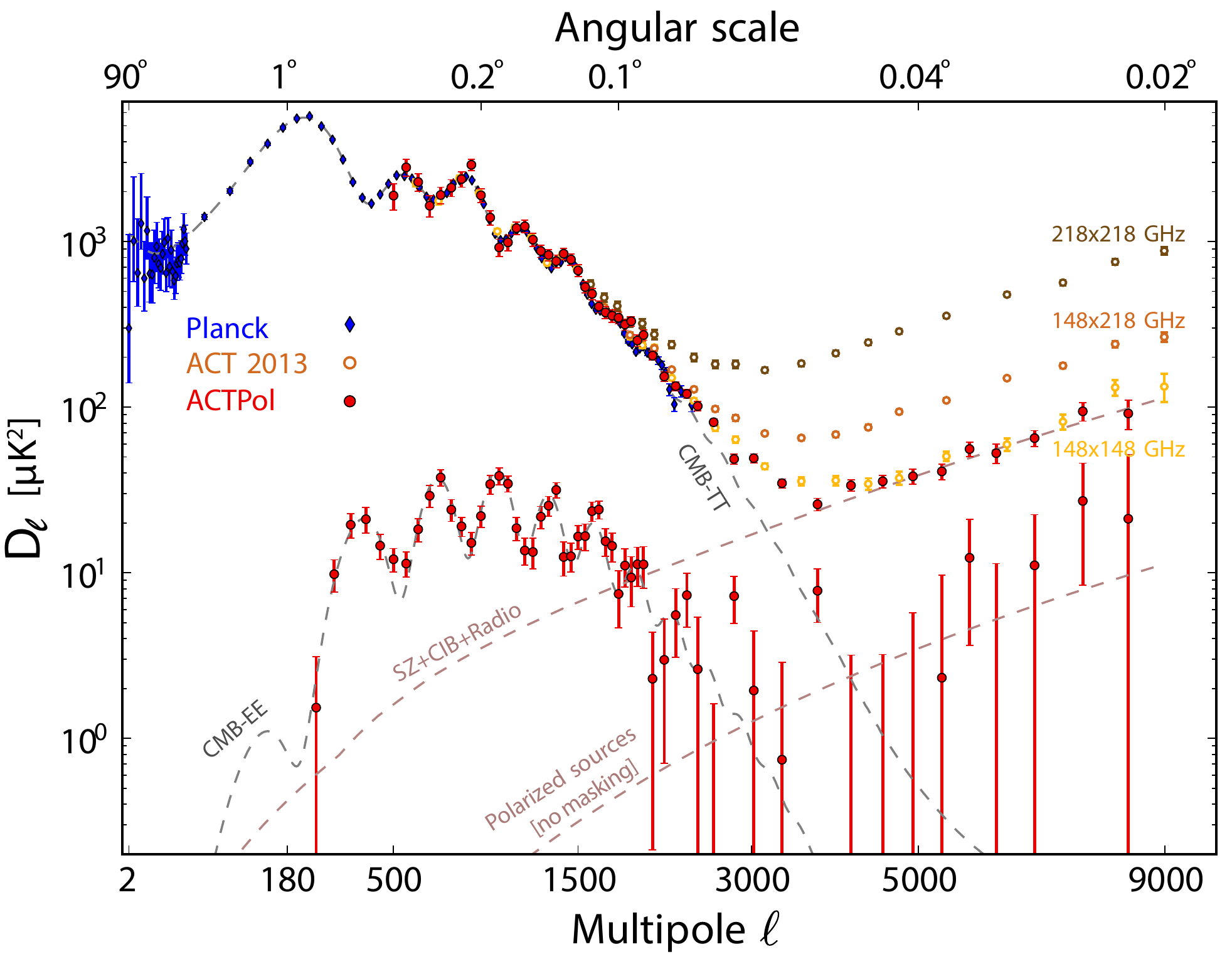}
\caption{ The {\it Planck}, ACT, and ACTPol data. Many {\it Planck} points for TT are obscured by the ACT data for $1000<\ell<2500$. The model spectra labeled CMB-TT and CMB-EE are for  `Planck+WP+highL' \citep[]{planck_params/2013}. 
It is clear that the same model is an excellent fit to the TT and EE data (see \S\ref{sec:LCDM}). Recently \map\ \citep{bennett/etal/2013} and SPT \citep{story/etal/2013} have also published new data on the TT spectrum in this range, which are not shown here. All measurements are broadly consistent.
The best-fitting Poisson polarized source level is shown, with no sources masked.
A non-zero level is preferred, but the distribution is consistent with zero at 95\% confidence, with $a_p^{\rm pol}<2.4$.
The $x$-axis is scaled as $\ell^{0.45}$ to emphasize the mid-$\ell$ range.}
\label{fig:tteespec}
\end{figure*}

\subsection{Foreground emission}
\label{subsec:fg}

We test for foreground emission in the temperature maps by correlating the ACTPol maps with the FDS dust template map \citep{finkbeiner/davis/schlegel/1999}. The dust level in this template has been shown to be consistent with the {\it Planck} 353~GHz maps \citep{planck_dust/2013} at the $30\%$ level. The predicted contribution of dust to the temperature anisotropy power spectrum is measurable but small, less than $2$ $\mu$K$^2$ at $\ell = 2000$ as shown in Figure \ref{fig:dust}. We do not correct for it in the maps or likelihood at this stage.

Based on the recent results from {\it Planck} \citep{planck_dust/2014}, the polarization fraction in D1, D5, and D6 is roughly 5\%. We take 10\% as an upper limit and thus the contribution from polarized dust emission to the power spectrum is expected to be less than $0.02$ $\mu$K$^2$ at $\ell=2000$. The contribution from polarized synchrotron emission is expected to be at this level or smaller. A full analysis must await the public release of the \planck\ polarization maps.  However, the consistency shown in Table \ref{tab:nulls} between patches D1, D5, and D6, each with different foreground levels, suggests that any possible contribution is small compared to the cosmological polarization signal.

\begin{figure*}
	\centering
	\includegraphics[width=0.8\textwidth]{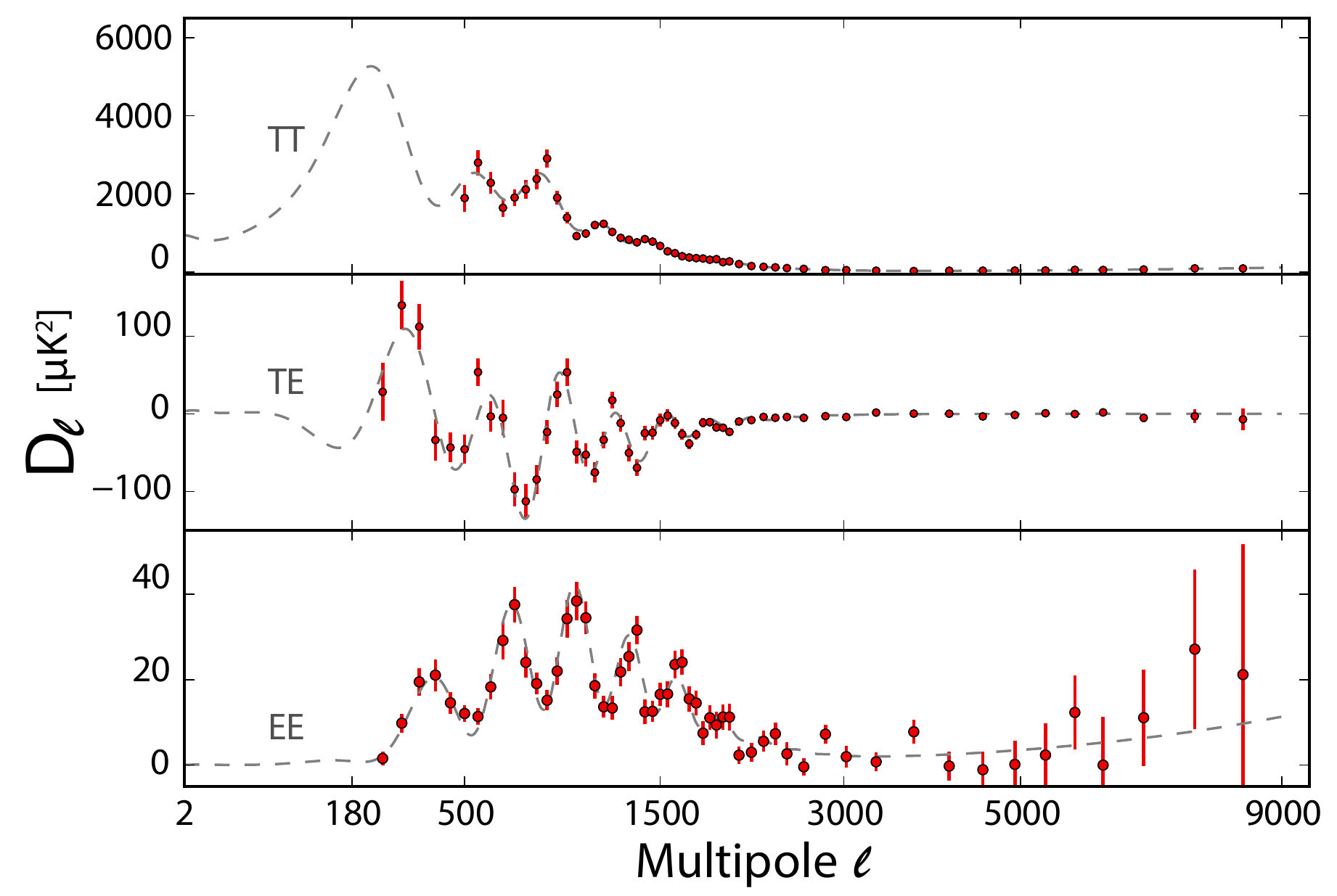}
\caption{The ACTPol TT, TE, and EE power spectra, together with the best-fitting \LCDM\ cosmological model and foreground components. Six acoustic peaks are seen  in the E-mode polarization, out of phase with the temperature peaks and with the TE correlation pattern predicted by the standard model.}
\label{fig:spectra}
\end{figure*}

\begin{figure}
	\includegraphics[width=\columnwidth]{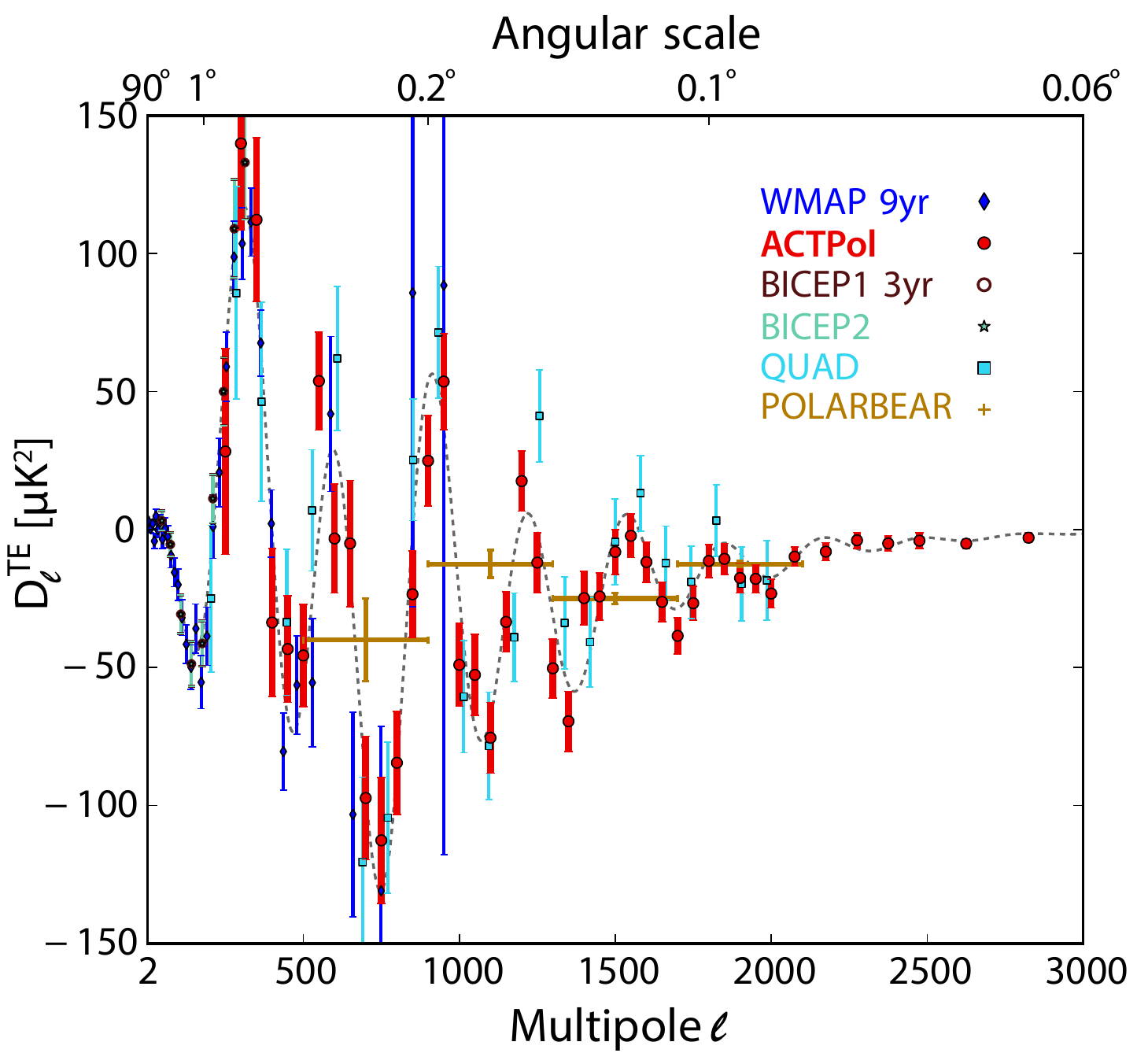}
	\caption{The ACTPol TE spectrum  together with results from \map\ 
		\citep{bennett/etal/2013}, QUAD \citep{brown/etal/2009}, BICEP1
		\citep{bicep/2013}, BICEP2 \citep{bicep2a/2014}, and \pbear{}
		\citep{pbear-eebb/2014}. For ACTPol we correlate with the ACTPol
		temperature maps, but we could reduce error bars by also
		correlating with {\it Planck} and/or ACT temperature maps. \planck{}
		has shown a plot of TE and EE \citep[Figure 11 of][]{planck_params/2013},
		but the data are not yet available.
}
\label{fig:tespec}
\end{figure}

\begin{figure}
	\includegraphics[width=\columnwidth]{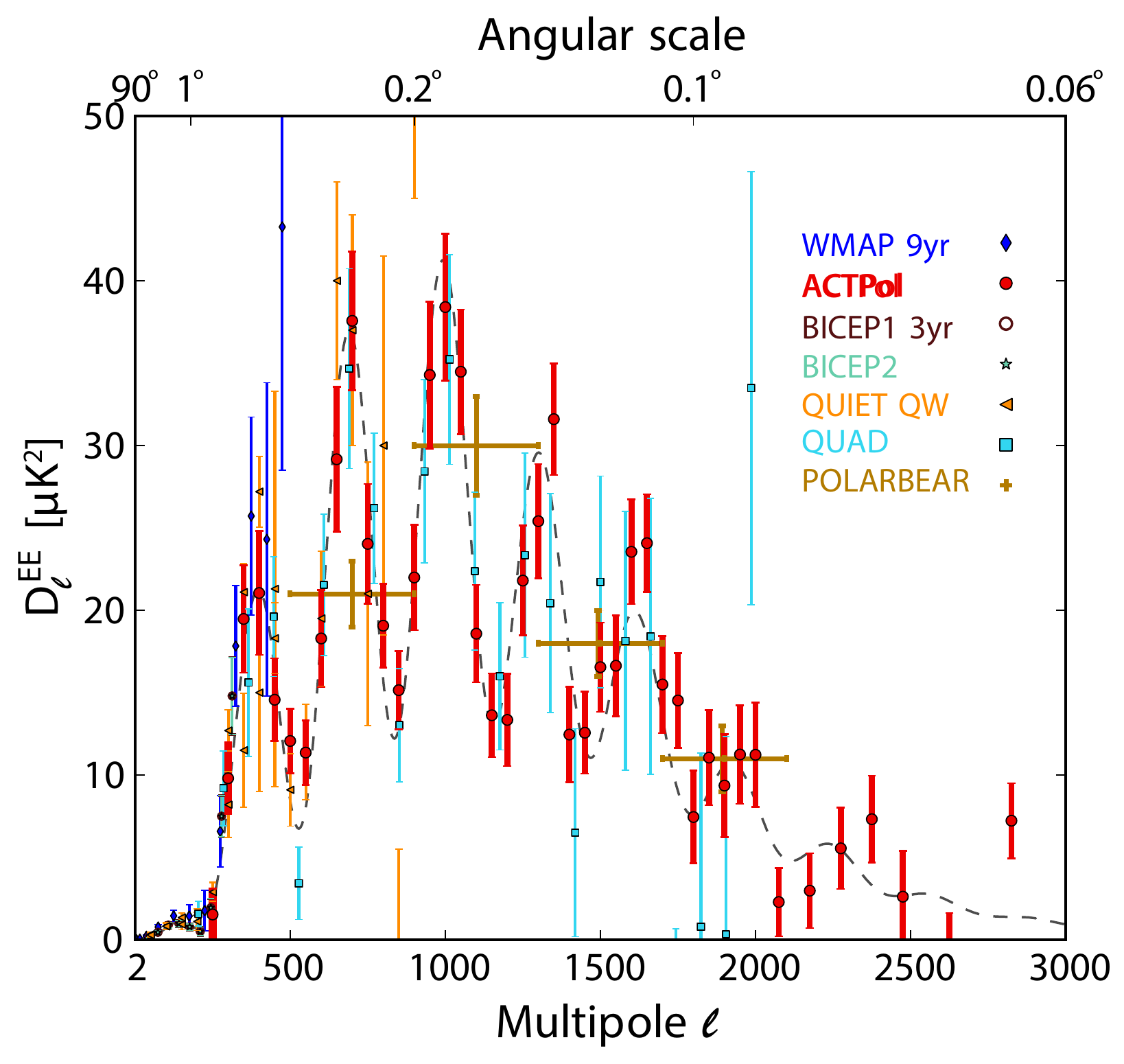}
\caption{The ACTPol data together with results from other EE measurements over the past five years, as in Figure \ref{fig:tespec}, and also including QUIET Q and W bands \citep{quiet-q/2011,quiet-w/2012}. 
}
\label{fig:eespec}
\end{figure}

\begin{figure}
	\includegraphics[width=\columnwidth]{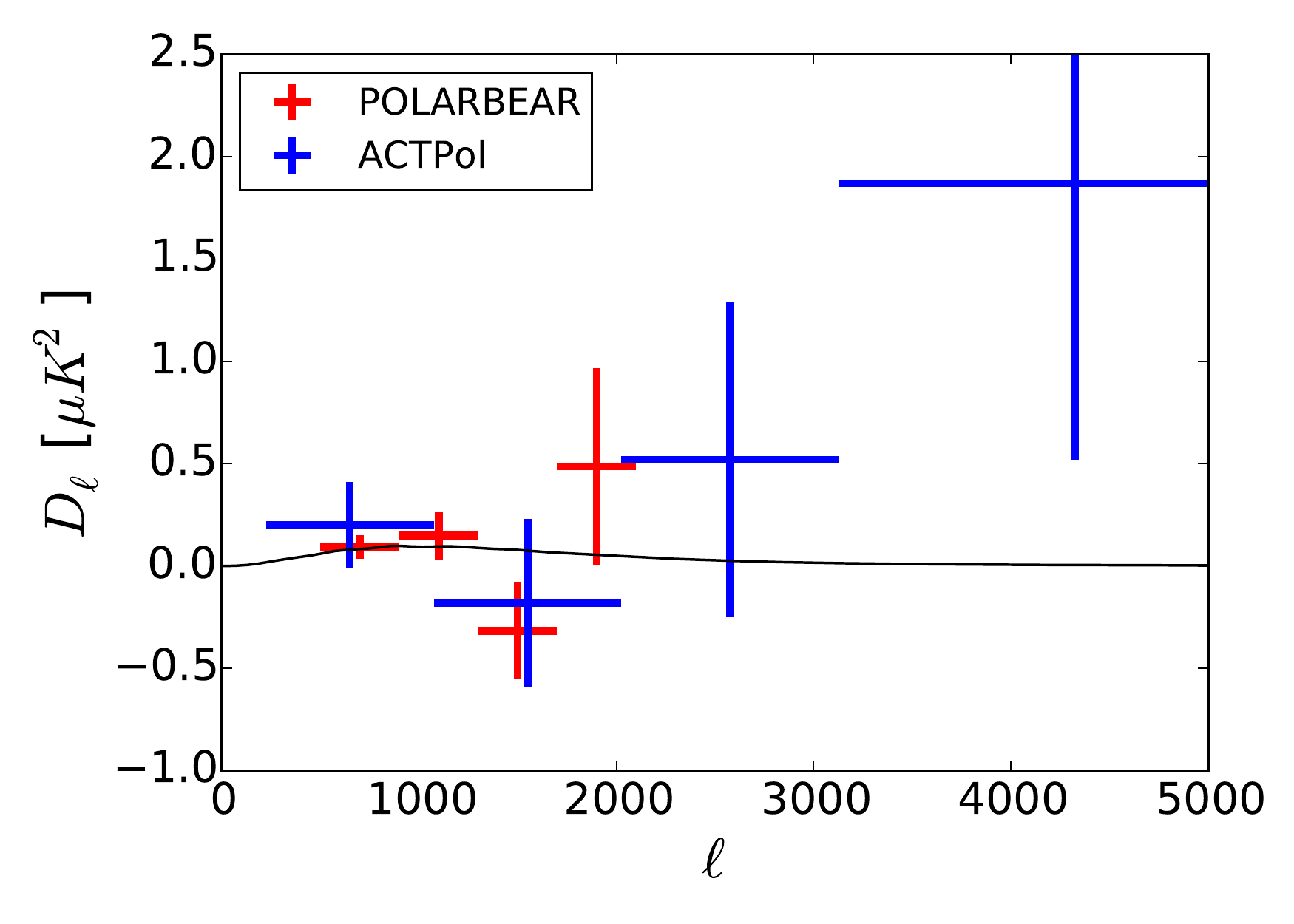}
	\caption{The ACTPol BB data rebinned into four bins, together with data from
	\citep{pbear-eebb/2014}. The black curve shows the expected power from lensing B-modes.
	The ACTPol data points are consistent with zero.}
	\label{fig:bbspec}
\end{figure}

\begin{table}
	\setlength{\tabcolsep}{4mm}
	\caption{Rebinned BB power spectrum, ${\cal{D}}_\ell = \ell(\ell+1)C_{\ell}/2\pi$ ($\mu$K$^2$), shown in Figure~\ref{fig:bbspec}.}
	\label{tab:bbspec}
	\vspace{-0.2in}
	\begin{center}
	\begin{tabular}{ c c  | r  r }
	\hline
	\hline
	$\ell$ & $\ell$ range & \multicolumn{2}{c}{BB} \\
  & & ${\cal{D}}_\ell$ & $\sigma ({\cal{D}}_\ell)$ \\
  \hline
	 650 & $ 225- 1075$ &  0.20 & 0.21 \\
	1550 & $1075- 2025$ & -0.18 & 0.41 \\
	2575 & $2025- 3125$ &  0.52 & 0.77 \\
	4325 & $3125- 5525$ &  1.87 & 1.35 \\
	\hline
\end{tabular}
\end{center}
\end{table}

\subsection{TT,TE,EE,BB}

Figure~\ref{fig:tteespec} shows the combined ACTPol TT and EE spectra along
with independent data sets from ACT temperature measurements and {\it Planck}. The full TT/TE/EE set is shown in Figure \ref{fig:spectra}. The TE spectrum is shown
in Figure~\ref{fig:tespec}, together with results from other CMB polarization
experiments. These and the spectra from Figure~\ref{fig:resids} are presented in
Table~\ref{tab:spectra}. The bandpowers are not significantly correlated, and
bandpower window functions are computed as in \citet{das/etal/2014} to compare
theory models  to these data. Figure~\ref{fig:eespec} shows the EE spectrum on
a linear scale, and Figure~\ref{fig:bbspec} shows the BB spectrum rebinned into 4 bins
based on the numbers in Table~\ref{tab:bbspec}. Rebinning to wider bins increases the
signal-to-noise ratio at the cost of resolution.

We observe six acoustic
peaks in the EE power spectrum, out of phase with the TT spectrum as expected
in the standard cosmological model, and six peak/troughs in the TE
cross-correlation. We detect no significant BB power.

As a simple test, we find the $\chi^2$ for the ACTPol EE data compared to the $\Lambda$CDM 6-parameter model using  a)  the \map{}+ACT parameters \citep{calabrese/etal/2013}, and b) the \planck\ best-fit parameters \citep{planck_params/2013}. The reduced $\chi^2$ values for the two models are 1.09 and 1.12 respectively
with 55 dof and no free parameters (with PTE of 0.30 and 0.25).  For the TE data the reduced $\chi^2$
for the two models are 1.26 and 1.24, again with 55 dof and no free parameters (PTE of 0.09 and 0.18).

\begin{table}
\caption{\small Comparison of cosmological Parameters and 68\% confidence intervals for different data sets.}
\begin{tabular}{ c  c  c  c }
\hline
\hline
  & \map{}+ACT\footnotemark[1]  &    Planck\footnotemark[2] & ACTPol \\%& ACTPol TE+EE \\
& & & TE,EE\footnotemark[3]\\
 \hline
$100\Omega_bh^2$  & $2.247  \pm 0.041$   & $2.207  \pm0.027$  & $2.073 \pm 0.135$ \\% & $0.02197\pm0.00026$ \\
$\Omega_ch^2$     & $0.1143 \pm 0.0044$  & $0.1198 \pm0.0026$ & $0.131 \pm 0.015$ \\%& $0.1203\pm0.0025$    \\
$10^4\theta_A$     & $103.95 \pm 0.19$    & $104.132\pm0.063$  & $104.12\pm 0.31$  \\%& $1.04106\pm0.00060$  \\
$\ln(10^{10}A_s)$ & $3.094  \pm 0.041$   & $3.090  \pm0.025$  & $3.190 \pm 0.085$ \\%& $3.087\pm0.025$  \\
$n_s$             & $0.970  \pm 0.011$   & $0.9585 \pm0.0070$ & [$0.970\pm0.011$] \\% &  $0.9563\pm0.0075$ \\
$\tau$            & $0.089  \pm 0.013$   & $0.091  \pm0.0135$ & [$0.089\pm0.013$] \\%&  $0.089\pm0.0130$ \\
\hline
Derived\footnotemark[4] &&&\\
$\sigma_8$        & $0.830  \pm 0.021$   & $0.828  \pm0.012$  & $0.933 \pm0.064$  \\%& $0.828\pm0.012$\\
$H_0$             & $69.7   \pm 2.0$     & $67.3   \pm1.2$    & $63.2  \pm5.5$    \\%& $67.0\pm1.1$\\
\hline
\end{tabular}
\footnotetext[1]{Joint analysis of \map{}+ACT as described in the text, assuming massless neutrinos.}
\footnotetext[2]{Parameters from `Planck+WP+highL' \citet{planck_params/2013}, assuming a $0.06$~eV summed neutrino mass.}
\footnotetext[3]{Parameters use just ACTPol TE and EE data, with priors imposed on $\tau$ and $n_s$ from \map{}+ACT (given in brackets) and assuming massless neutrinos.}
\footnotetext[4]{The derived parameters $\sigma_8$ and $H_0$ (in units of km s$^{-1}$ Mpc$^{-1}$) are also presented.}
\label{tab:params}
\end{table}

\subsubsection{\LCDM}
\label{sec:LCDM}
Another test of the standard \LCDM\ cosmological model is to fit its parameters from just the ACTPol EE and TE data. In this combination, \map{} is used to put a prior on the optical depth and scalar spectral index, since ACTPol does not measure the largest angular scales. We estimate parameters using standard methods as in \citet{sievers/etal/2013} and \citet{calabrese/etal/2013}, and marginalize over Poisson source powers for the TE and EE spectra. Other foregrounds are assumed to be unpolarized.

\begin{figure}
	\includegraphics[width=\columnwidth]{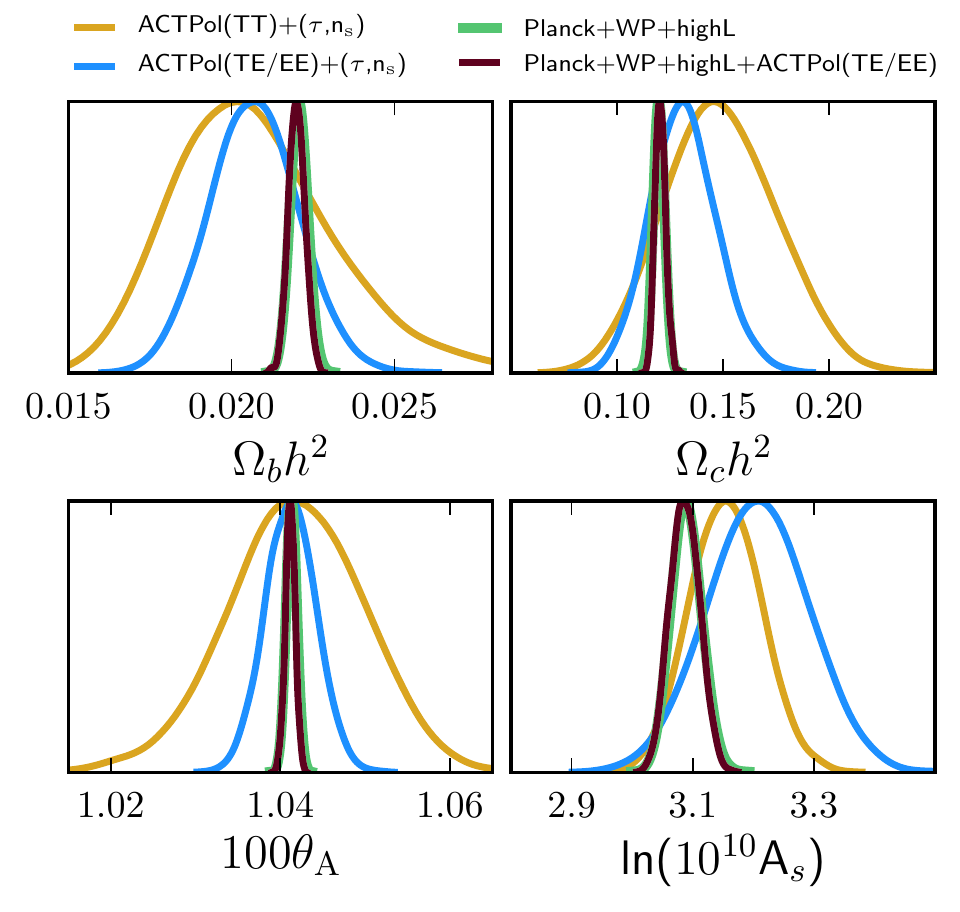}
	\caption{\LCDM\ parameters estimated from the ACTPol TE and EE data alone(with a prior on the optical depth and spectral index from \map), and ACTPol TT alone. They are compared to the constraints from {\it Planck} temperature data, and combined \planck\ and ACTPol TE and EE. The temperature and polarization data give consistent results.}
	\label{fig:params}
\end{figure}

The results are reported in Table~\ref{tab:params} and shown in Figure \ref{fig:params} for the physical baryon density, $\Omega_bh^2$, the physical cold dark matter density, $\Omega_ch^2$, the acoustic scale $\theta_A$, and the amplitude of primordial curvature perturbations, $A_s$, defined at pivot scale $k_0=0.05$~Mpc$^{-1}$. The polarization data are in excellent agreement with the standard model constrained by the {\it Planck} temperature data \citep{planck_params/2013}.

We repeat the same test with just the ACTPol TT data, including the foreground model as in \citet{dunkley/etal/2013} to account for Poisson and clustered point sources and the thermal and kinematic Sunyaev-Zel'dovich effects. The parameters are consistent with the ACTPol TE/EE results, but more weakly constrain the acoustic scale and the physical baryon and CDM densities (see Figure~\ref{fig:params}). This highlights the potential of the E-mode polarization signal for cosmological constraints \citep[see e.g.,][]{galli/etal/2014}, with sharper acoustic features and less contamination from atmosphere and foregrounds.
At this stage of measurement, however, the polarization data are not as precise as the {\it Planck} or \map{} temperature data primarily because they are taken over a relatively small region of sky. 
Parameter constraints from ACTPol combined with {\it Planck} are currently dominated by the temperature data.

\subsubsection{Peak/Dip Phases}
\label{sec:phases}

\LCDM\ predicts that the TT and polarization peaks should be out of phase.  We
test this quantitatively, in a manner similar to that used
in \citet{readhead/etal/2004}. We start by computing the theoretical
\LCDM{} TE and EE power spectra based on the best-fit parameters from Section~\ref{sec:LCDM}.
Since we wish to test for any unexpected phase shift between the TE and EE spectra,
we construct a simple parametric model that approximates the \LCDM{} spectra,
but has individually adjustable phases for each spectrum. This model takes the
form $r_1(\ell)+r_2(\ell)\cos(2\pi \ell/L + \phi)$
where $r_1(\ell)$ and $r_2(\ell)$ are rational functions with third-order
polynomials in both the numerator and denominator, $L$ is the period
of the peaks, and $\phi$ is the phase of the pattern, all of which are fit
independently to each spectrum such that the deviation from the \LCDM{}
spectra is minimized in the range $100<\ell<2000$. The result is
best-fit rational functions $\hat r_1(\ell)$ and $\hat r_2(\ell)$ and a best-fit 
phase parameter $\hat\phi$ for each of TT and EE. These modulated
rational function models are very good fits to the \LCDM{} spectra,
but there is still a small residual. We therefore make the replacement
$\hat r_1(\ell) \rightarrow \hat r_1(\ell) + \textrm{residual}(\ell)$, such that
at $\phi=\hat\phi$  the model exactly reproduces the \LCDM{} spectra.

With these models in hand, we are now in the position to ask whether
our observed power spectra prefer the same $\phi$ values as \LCDM{} does.
For each of TE and EE, we fit linear a three-parameter model
$a_1 \hat r_1(\ell) + a_2 \hat r_2(\ell)\cos(2\pi\ell/L+\hat\phi)+a_3 \hat r_2(\ell)\sin(2\pi\ell/L+\hat\phi)$
to our observations in the range $225<\ell<2000$.
This model effectively encodes a phase shift $\Delta\phi = \phi_\textrm{data}-\hat\phi = {\rm Arg}(a_2+ia_3)$
(our main interest here) as well as a wave amplitude and an
overall amplitude factor. 

\begin{figure}
	\includegraphics[width=\columnwidth]{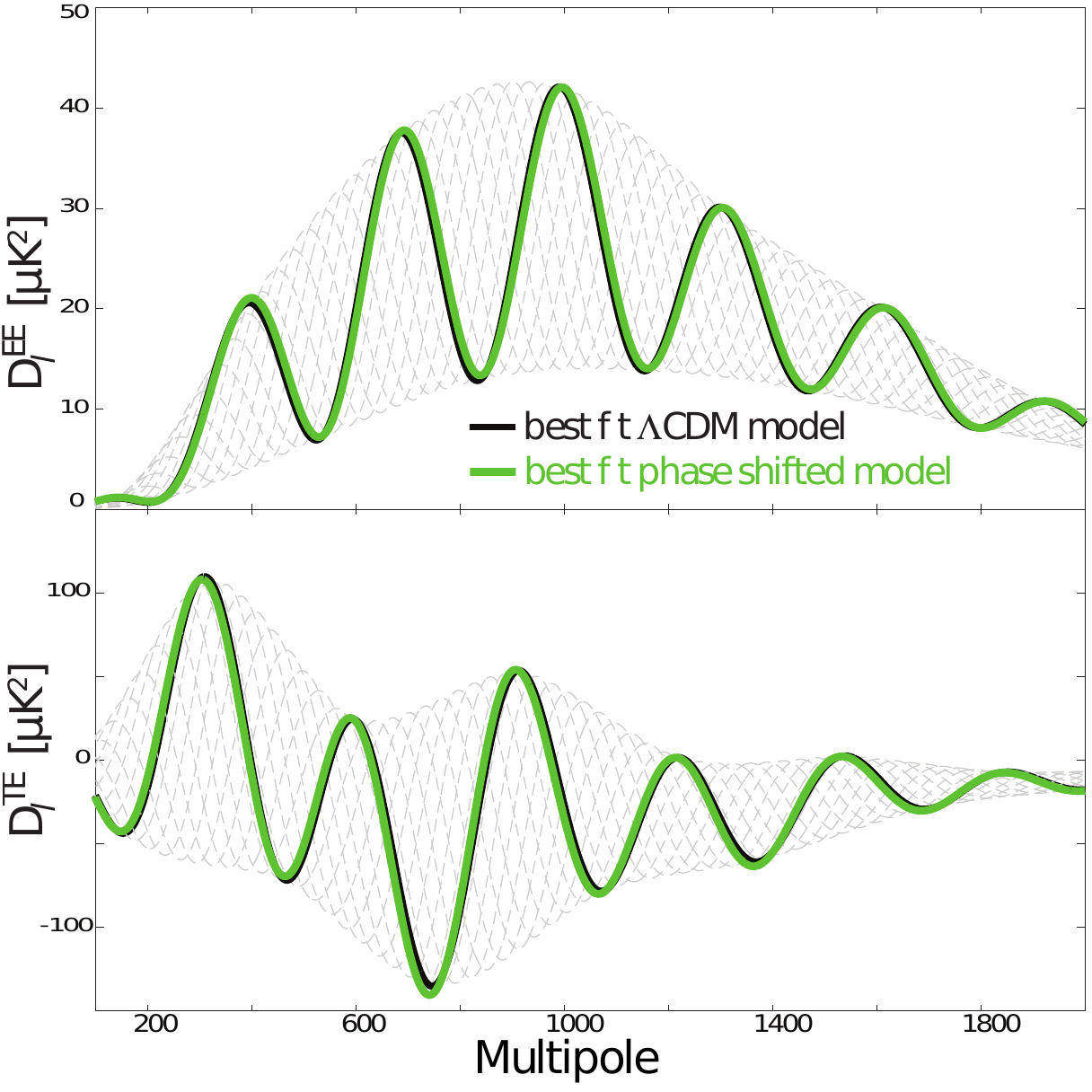}
\caption{Recovered amplitudes and phases of the polarization peak/dip pattern from
ACTPol data.  The black curves mark the best-fit \LCDM\ model power
spectrum.  The green curves show the best-fit amplitude/phase EE (top) and TE (bottom) models to
the ACTPol data as described in the text.  The thin lines in
the background show the envelope of the phase-shifted model.  
 The polarization data are in excellent agreement with the \LCDM\
 prediction.}
\label{fig:phases}
\end{figure}

The resulting fits can be seen
in Table~\ref{tab:phase}, and are consistent with the \LCDM{} expectations (i.e.
the phase shifts are all consistent with zero). A graphical illustration of the
fit compared to the prediction and model space can be seen in Figure~\ref{fig:phases}.
\begin{table}
	\caption{Results of fitting a phase shift in the observed TE and EE spectra
	relative to the \LCDM{} best-fit model.}
	\begin{tabular}{lrrr}
		& TE & EE & TE+EE \\
		\hline
		\hline
		$a_1$        &$ 1.036\pm0.066$ &$1.008\pm0.032$ &$1.014\pm0.032$ \\
		$a_2$        &$ 1.000\pm0.080$ &$0.985\pm0.088$ &$0.986\pm0.061$ \\
		$a_3$        &$-0.108\pm0.080$ &$0.108\pm0.088$ &$0.003\pm0.061$ \\
		\hline
		$\phi_\textrm{data}$ (${}^\circ$)       &$23.0$           &$-64.9$         & \\
		$\Delta\phi$ (${}^\circ$) &$-6.2\pm4.6$     &$6.2\pm5.1$     &$0.2\pm3.6$ \\
		\hline
	\end{tabular}
\textsc{Note:} In the last row, a single common phase shift is fit
	jointly for TE and EE while still using their individual $\phi$
        angles. The fits
	are in agreement with the \LCDM{} expectations.
	\label{tab:phase}
\end{table}

\pagebreak

\subsubsection{Polarized point sources}

\begin{figure}
	\centering
	\setlength{\tabcolsep}{1.5pt}
	\begin{tabular}{m{3mm}m{13mm}m{13mm}m{13mm}m{13mm}m{13mm}m{13mm}}
		{\bf T} &
		\includegraphics[width=0.15\columnwidth]{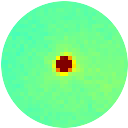} &
		\includegraphics[width=0.15\columnwidth]{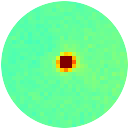} &
		\includegraphics[width=0.15\columnwidth]{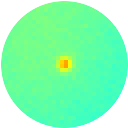} &
		\includegraphics[width=0.15\columnwidth]{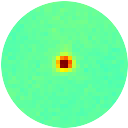} &
		\includegraphics[width=0.15\columnwidth]{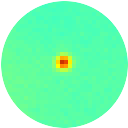} &
		\includegraphics[width=0.15\columnwidth]{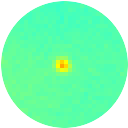} \\
		{\bf Q} &
		\includegraphics[width=0.15\columnwidth]{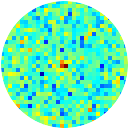} &
		\includegraphics[width=0.15\columnwidth]{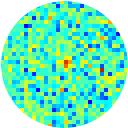} &
		\includegraphics[width=0.15\columnwidth]{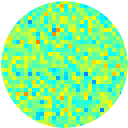} &
		\includegraphics[width=0.15\columnwidth]{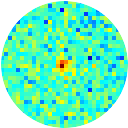} &
		\includegraphics[width=0.15\columnwidth]{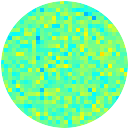} &
		\includegraphics[width=0.15\columnwidth]{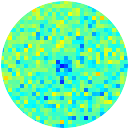} \\
		{\bf U} &
		\includegraphics[width=0.15\columnwidth]{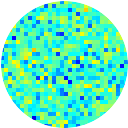} &
		\includegraphics[width=0.15\columnwidth]{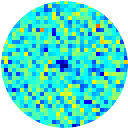} &
		\includegraphics[width=0.15\columnwidth]{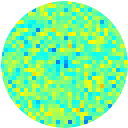} &
		\includegraphics[width=0.15\columnwidth]{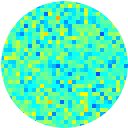} &
		\includegraphics[width=0.15\columnwidth]{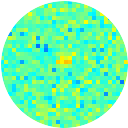} &
		\includegraphics[width=0.15\columnwidth]{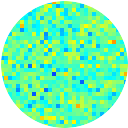}
	\end{tabular}
	\caption{Three of the highest signal-to-noise polarized point sources from each of patch D5
		(left) and D6 (right).
		Each disk has a radius of 8', with the value range being $\pm 2000\mu$K
		for T and $\pm 200\mu$K for Q and U.
We do not mask polarized sources in this analysis. The sources may be associated with
(from left to right) [HB89]~2332-017, [HB89]~2335-027,
SDSS~J001130.40+005751.7, PKS~0214-085, [HB89]~0226-038 and
	PKS~0205-010.
}
	\label{fig:ptsrc}
\end{figure}

We do not detect significantly polarized point sources in the ACTPol maps. Six of the highest signal-to-noise sources found in the D5 and D6 patches are shown in Figure \ref{fig:ptsrc}; the polarization is barely detectable. For the temperature spectra, point sources above 15 mJy are masked, but no sources exceed this threshold in polarization and none are masked.

We model the Poisson tail of the temperature spectrum as
\begin{equation}
{\cal D}_\ell = (a_s+a_d)\bigg(\frac{\ell}{3000}\bigg)^2 \mu{\rm K}^2
\end{equation}
\citep{dunkley/etal/2011}, where $a_s$ is the amplitude for the residual
unmasked radio/synchrotron sources and $a_d$ is the amplitude for the pervasive
dusty star forming galaxy (DSFG) or CIB component. The latter is unresolved.
The two components are separated in ACT temperature measurements with two
observing frequencies. In ACTPol we currently have just one frequency and so
place a limit on the combined Poisson power: $a_p=a_s+a_d$. In
\citet{sievers/etal/2013} we found $a_s=3.1\pm0.4$ and
$a_d=7.0\pm0.5$, for a total of $a_p=10.3\pm0.6$ for
the TT data. With the ACTPol TT data we find a consistent level of
$a_p=10.9\pm1.5$, for the same masking threshold.

Without masking any point sources in the EE data we find $a_p^{\rm
pol}=1.5\pm0.6$ at 68\% confidence, or $a_p^{\rm
pol}<2.4$ at 95\% confidence. In flux units this corresponds to
$C_\ell = 0.15^{+0.05}_{-0.07}~{\rm Jy}^2/{\rm sr}$, or $<0.24~{\rm Jy}^2/{\rm
sr}$ (95\% CL), at 146~GHz, and puts a limit on all polarized sources before
masking.

\section{Conclusions}
\label{sec:conc}

The polarization capabilities of ACTPol have already enabled new probes of the cosmological model. We have shown that ACT is capable of measuring polarization to high accuracy and of measuring CMB temperature and polarization during the day. With one third of the full complement of detectors observing at night over just 90 days, we have already made some of the most competitive measurements yet of CMB polarization at $\ell>1000$. The ACTPol EE, TE, TB, and BB data obtained to date are all in agreement with the standard model of cosmology. We anticipate substantial improvements with more detectors and observing time.

\acknowledgments

This work was supported by the U.S. National Science Foundation through awards
AST-0408698 and AST-0965625 for the ACT project, as well as awards PHY-0855887
and PHY-1214379. Funding was also provided by Princeton University, the
University of Pennsylvania, and a Canada Foundation for Innovation (CFI) award
to UBC. ACT operates in the Parque Astron\'omico Atacama in northern Chile
under the auspices of the Comisi\'on Nacional de Investigaci\'on Cient\'ifica y
Tecnol\'ogica de Chile (CONICYT). Computations were performed on the GPC
supercomputer at the SciNet HPC Consortium. SciNet is funded by the CFI under
the auspices of Compute Canada, the Government of Ontario, the Ontario Research
Fund -- Research Excellence; and the University of Toronto. The development of
multichroic detectors and lenses was supported by NASA grants NNX13AE56G and
NNX14AB58G. CM acknowledges support from NASA grant NNX12AM32H. Funding from
ERC grant 259505 supports SN, JD, EC, and TL. HT is supported by grants NASA
ATP NNX14AB57G, DOE DE-SC0011114, and NSF AST-1312991.
BS, BK, CM, and EG are funded by NASA Space Technology Research Fellowships.
R.D received funding from the CONICYT grants QUIMAL-120001 and FONDECYT-1141113.
We thank our many
colleagues from ABS, ALMA, APEX, and \pbear{} who have helped us at critical
junctures. Colleagues at AstroNorte and RadioSky provide logistical
support and keep operations in Chile running smoothly.
We thank Jesse Treu for multiple suggestions and comments.
We also thank the Mishrahi Fund and the Wilkinson Fund for their generous
support of the project.

\begin{table*}
\caption{Combined Power Spectra, ${\cal{D}}_\ell = \ell(\ell+1)C_{\ell}/2\pi$ ($\mu$K$^2$), shown in Figures~\ref{fig:resids} and \ref{fig:spectra}.\protect\footnotemark}
\vspace{-0.2in}
\begin{center}
\begin{tabular}{ c c  | r  r |  r  r | r  r | r  r | r  r | r  r }
\hline
\hline
$\ell$ & $\ell$ range & \multicolumn{2}{c|}{TT} & \multicolumn{2}{c|}{TE} & \multicolumn{2}{c|}{EE} & \multicolumn{2}{c|}{BB} & \multicolumn{2}{c|}{TB} & \multicolumn{2}{c}{EB} \\

  & & ${\cal{D}}_\ell$ & $\sigma ({\cal{D}}_\ell)$
& ${\cal{D}}_\ell$ & $\sigma ({\cal{D}}_\ell)$
& ${\cal{D}}_\ell$ & $\sigma ({\cal{D}}_\ell)$
& ${\cal{D}}_\ell$ & $\sigma ({\cal{D}}_\ell)$
& ${\cal{D}}_\ell$ & $\sigma ({\cal{D}}_\ell)$
& ${\cal{D}}_\ell$ & $\sigma ({\cal{D}}_\ell)$
\\
\hline
 250 & $ 225- 275$ & 4049.4 & 1476.4 &   28.3 &   37.3 &    1.5 &    1.6 &    0.1 &    1.2 &   25.1 &   25.9 &    0.8 &    0.9 \\
 300 & $ 275- 325$ & 3366.7 &  788.5 &  139.9 &   31.2 &    9.8 &    2.2 &    1.4 &    0.8 &    8.5 &   17.3 &    1.2 &    0.9 \\
 350 & $ 325- 375$ & 2377.3 &  510.0 &  112.2 &   29.7 &   19.5 &    3.3 &    0.3 &    0.7 &   -5.7 &   12.4 &    0.6 &    0.9 \\
 400 & $ 375- 425$ & 1366.2 &  376.3 &  -33.7 &   26.8 &   21.0 &    3.8 &    1.0 &    0.7 &    9.3 &   11.4 &    0.5 &    1.1 \\
 450 & $ 425- 475$ & 1683.2 &  310.0 &  -43.3 &   19.4 &   14.6 &    2.5 &    1.1 &    0.7 &   10.9 &    9.2 &    0.2 &    0.9 \\
 500 & $ 475- 525$ & 1892.9 &  342.9 &  -45.6 &   18.5 &   12.1 &    2.0 &   -0.2 &    0.7 &   23.4 &   10.4 &    0.1 &    0.8 \\
 550 & $ 525- 575$ & 2801.0 &  323.5 &   53.8 &   17.8 &   11.4 &    2.0 &    0.3 &    0.7 &  -14.1 &    9.8 &   -0.7 &    0.8 \\
 600 & $ 575- 625$ & 2284.9 &  278.6 &   -3.3 &   19.8 &   18.3 &    3.0 &   -0.3 &    0.8 &  -11.5 &    9.1 &   -0.1 &    1.0 \\
 650 & $ 625- 675$ & 1644.6 &  242.7 &   -5.1 &   22.8 &   29.2 &    4.4 &   -0.7 &    0.8 &    4.9 &    9.3 &   -0.1 &    1.2 \\
 700 & $ 675- 725$ & 1907.0 &  218.9 &  -97.3 &   22.4 &   37.6 &    4.2 &   -0.2 &    0.8 &   -4.8 &    8.6 &    1.2 &    1.2 \\
 750 & $ 725- 775$ & 2112.9 &  246.5 & -112.6 &   22.8 &   24.0 &    3.6 &    0.6 &    0.8 &   -2.4 &    8.8 &   -0.3 &    1.1 \\
 800 & $ 775- 825$ & 2381.3 &  257.2 &  -84.5 &   18.7 &   19.1 &    2.5 &   -1.0 &    1.0 &   -4.9 &   10.0 &    0.7 &    1.0 \\
 850 & $ 825- 875$ & 2904.4 &  228.6 &  -23.5 &   15.7 &   15.2 &    2.4 &    0.3 &    1.0 &    4.2 &    9.1 &    1.1 &    1.0 \\
 900 & $ 875- 925$ & 1901.5 &  181.3 &   24.9 &   16.4 &   22.0 &    3.2 &    1.1 &    1.0 &   13.0 &    8.3 &    0.1 &    1.2 \\
 950 & $ 925- 975$ & 1393.4 &  140.3 &   53.6 &   17.4 &   34.3 &    4.5 &   -1.0 &    1.2 &   -9.0 &    8.2 &    1.8 &    1.5 \\
1000 & $ 975-1025$ &  920.8 &  109.2 &  -49.0 &   15.0 &   38.4 &    4.5 &   -1.4 &    1.1 &    8.8 &    7.1 &   -0.1 &    1.5 \\
1050 & $1025-1075$ &  987.4 &  107.8 &  -52.7 &   14.7 &   34.5 &    3.8 &   -0.4 &    1.4 &   -4.3 &    7.5 &    1.5 &    1.4 \\
1100 & $1075-1125$ & 1203.5 &  105.2 &  -75.5 &   12.9 &   18.6 &    3.0 &   -0.0 &    1.3 &  -17.4 &    7.6 &    0.6 &    1.3 \\
1150 & $1125-1175$ & 1234.2 &  107.1 &  -33.5 &   10.9 &   13.6 &    2.5 &   -0.9 &    1.4 &    9.4 &    7.6 &    1.3 &    1.3 \\
1200 & $1175-1225$ & 1026.2 &   90.6 &   17.6 &   10.9 &   13.3 &    2.8 &    0.6 &    1.5 &   -2.8 &    7.2 &    0.7 &    1.5 \\
1250 & $1225-1275$ &  875.3 &   74.9 &  -12.0 &   10.9 &   21.8 &    3.3 &   -1.4 &    1.5 &    4.8 &    6.9 &   -1.8 &    1.5 \\
1300 & $1275-1325$ &  828.4 &   65.6 &  -50.3 &   10.7 &   25.4 &    3.5 &   -0.7 &    1.5 &    8.6 &    6.3 &   -0.8 &    1.5 \\
1350 & $1325-1375$ &  761.1 &   67.4 &  -69.5 &   10.9 &   31.6 &    3.4 &    2.0 &    1.7 &   -7.0 &    6.8 &    1.5 &    1.7 \\
1400 & $1375-1425$ &  843.5 &   66.0 &  -24.9 &    9.7 &   12.5 &    2.9 &    1.2 &    1.7 &   -6.0 &    6.5 &   -1.2 &    1.5 \\
1450 & $1425-1475$ &  778.9 &   60.2 &  -24.3 &    8.5 &   12.6 &    2.5 &   -2.3 &    1.7 &   -5.9 &    6.6 &    0.7 &    1.4 \\
1500 & $1475-1525$ &  668.3 &   54.3 &   -8.1 &    8.1 &   16.5 &    2.7 &   -1.2 &    1.7 &  -13.7 &    6.2 &   -0.5 &    1.5 \\
1550 & $1525-1575$ &  530.9 &   42.9 &   -2.3 &    7.9 &   16.6 &    3.1 &   -1.5 &    2.0 &   -3.2 &    5.8 &   -1.0 &    1.7 \\
1600 & $1575-1625$ &  482.7 &   37.1 &  -11.8 &    7.3 &   23.5 &    3.2 &   -0.4 &    1.9 &   -5.5 &    5.3 &   -1.8 &    1.7 \\
1650 & $1625-1675$ &  404.7 &   33.2 &  -26.2 &    7.2 &   24.1 &    3.0 &   -0.6 &    2.1 &   -5.5 &    5.2 &   -0.5 &    1.7 \\
1700 & $1675-1725$ &  372.8 &   31.6 &  -38.5 &    6.6 &   15.5 &    2.9 &   -0.7 &    2.0 &   -2.8 &    5.1 &   -1.5 &    1.7 \\
1750 & $1725-1775$ &  356.3 &   31.0 &  -26.7 &    6.2 &   14.5 &    2.9 &    2.7 &    2.3 &    9.3 &    5.4 &   -0.5 &    1.7 \\
1800 & $1775-1825$ &  346.6 &   28.8 &  -11.4 &    6.1 &    7.5 &    2.8 &   -1.3 &    2.2 &    0.7 &    5.1 &   -1.3 &    1.7 \\
1850 & $1825-1875$ &  315.5 &   24.4 &  -10.6 &    5.6 &   11.1 &    2.9 &    0.4 &    2.4 &   -0.9 &    4.7 &    0.2 &    1.8 \\
1900 & $1875-1925$ &  330.0 &   22.2 &  -17.6 &    5.4 &    9.4 &    3.1 &   -0.7 &    2.5 &    3.8 &    4.7 &    0.9 &    1.9 \\
1950 & $1925-1975$ &  252.6 &   20.4 &  -17.9 &    5.1 &   11.2 &    3.0 &    4.8 &    2.5 &   -0.4 &    4.4 &   -0.6 &    1.8 \\
2000 & $1975-2025$ &  272.6 &   19.1 &  -23.2 &    5.2 &   11.2 &    3.2 &    0.2 &    2.7 &   -2.3 &    4.6 &   -0.9 &    2.0 \\
2075 & $2025-2125$ &  204.5 &   12.5 &   -9.8 &    3.5 &    2.3 &    2.1 &    0.8 &    1.9 &   -0.7 &    3.1 &    0.3 &    1.4 \\
2175 & $2125-2225$ &  153.2 &   10.3 &   -8.0 &    3.3 &    3.0 &    2.3 &   -2.5 &    2.1 &   -1.5 &    2.9 &    2.1 &    1.5 \\
2275 & $2225-2325$ &  133.9 &    8.3 &   -3.9 &    2.9 &    5.6 &    2.5 &   -1.5 &    2.2 &    1.1 &    2.8 &    0.3 &    1.6 \\
2375 & $2325-2425$ &  120.1 &    7.4 &   -5.1 &    2.9 &    7.3 &    2.6 &    0.6 &    2.3 &   -2.6 &    2.7 &    1.0 &    1.7 \\
2475 & $2425-2525$ &  101.4 &    6.7 &   -4.0 &    2.8 &    2.6 &    2.8 &    3.1 &    2.5 &    2.7 &    2.6 &    1.3 &    1.8 \\
2625 & $2525-2725$ &   81.1 &    4.1 &   -5.1 &    1.8 &   -0.4 &    2.1 &    4.3 &    1.9 &   -0.4 &    1.7 &    1.3 &    1.3 \\
2825 & $2725-2925$ &   48.6 &    3.5 &   -3.0 &    1.8 &    7.2 &    2.3 &   -1.3 &    2.2 &    0.9 &    1.7 &   -0.0 &    1.5 \\
3025 & $2925-3125$ &   49.2 &    3.1 &   -4.0 &    1.8 &    1.9 &    2.5 &    0.1 &    2.5 &   -0.3 &    1.7 &    0.1 &    1.7 \\
3325 & $3125-3525$ &   34.7 &    2.2 &    1.7 &    1.3 &    0.7 &    2.1 &    0.5 &    2.1 &   -0.5 &    1.3 &    0.3 &    1.5 \\
3725 & $3525-3925$ &   25.9 &    2.3 &    0.3 &    1.6 &    7.8 &    2.8 &    3.2 &    2.7 &   -0.1 &    1.5 &    0.6 &    1.8 \\
4125 & $3925-4325$ &   33.7 &    2.6 &    0.2 &    1.8 &   -0.3 &    3.4 &    2.3 &    3.4 &    0.6 &    1.8 &    0.3 &    2.3 \\
4525 & $4325-4725$ &   35.7 &    3.2 &   -3.3 &    2.3 &   -1.1 &    4.3 &    1.1 &    4.5 &    3.0 &    2.3 &    1.0 &    2.9 \\
4925 & $4725-5125$ &   38.2 &    3.9 &   -1.4 &    2.9 &    0.1 &    5.6 &   -0.2 &    5.6 &    0.7 &    2.8 &    2.1 &    3.9 \\
5325 & $5125-5525$ &   40.9 &    4.5 &    0.8 &    3.5 &    2.3 &    7.4 &   11.7 &    7.1 &   -6.7 &    3.5 &   -7.0 &    4.9 \\
5725 & $5525-5925$ &   55.8 &    5.7 &   -0.2 &    4.3 &   12.3 &    8.7 &   24.4 &    8.9 &    1.6 &    4.2 &   -8.7 &    6.0 \\
6125 & $5925-6325$ &   52.7 &    6.9 &    1.8 &    5.6 &   -0.0 &   11.4 &    2.6 &   11.2 &   -0.5 &    5.3 &    1.5 &    7.7 \\
6725 & $6325-7125$ &   65.0 &    7.2 &   -5.1 &    5.3 &   11.1 &   11.3 &   12.6 &   11.3 &    4.0 &    5.4 &   11.9 &    7.7 \\
7525 & $7125-7925$ &   94.3 &   12.0 &   -3.0 &    8.7 &   27.1 &   18.7 &   43.0 &   17.9 &   -2.9 &    8.6 &   25.5 &   12.7 \\
8325 & $7925-8725$ &   91.6 &   18.7 &   -7.0 &   14.0 &   21.2 &   30.6 &   18.6 &   30.3 &   14.4 &   14.0 &   14.1 &   20.8 \\

\hline
\end{tabular}
\footnotetext[1]{The overall temperature map calibration error is 2\%.}
\end{center}
\label{tab:spectra}
\end{table*}

\bibliographystyle{act}

\end{document}